\documentclass[a4paper,usenatbib]{mnras}

\usepackage{amsmath}
\usepackage{amssymb}
\usepackage{newtxtext,newtxmath}
\usepackage[T1]{fontenc}
\usepackage{ae,aecompl}
\usepackage{graphicx}
\usepackage{times}
\usepackage{xcolor}
\newcommand{\txd}{{\text{d}}}
\newcommand{\calE}{{\mathcal{E}}}
\newcommand{\calN}{{\mathcal{N}}}
\newcommand{\rb}{r_{\text{b}}}
\newcommand{\Psib}{\Psi_{\text{b}}}
\newcommand{\rref}{r_{\text{ref}}}

\title[Broken power-law and double power-law models]%
{The dynamical structure of broken power-law and double power-law models for dark matter haloes}

\author[M.~Baes \& P.~Camps]{Maarten Baes and Peter Camps\\%
Sterrenkundig Observatorium, Universiteit Gent, Krijgslaan 281 S9, B-9000 Gent, Belgium}

\date{Accepted 2021 March 01. Received 2021 February 26; in original form 2021 January 14.}

\begin{document}
\label{firstpage}
\pagerange{\pageref{firstpage}--\pageref{lastpage}}

\maketitle

\begin{abstract}
Galaxy kinematics and gravitational lensing are two complementary ways to constrain the distribution of dark matter on galaxy scales. The typical dark matter density profiles adopted in dynamical studies cannot easily be adopted in lensing studies. Ideally, a mass model should be used that has the global characteristics of realistic dark matter distributions, and that allows for an analytical calculation of the magnifications and deflection angles. A simple model with these properties, the broken-power-law (BPL) model, has very recently been introduced. We examine the dynamical structure of the family of BPL models. We derive simple closed expressions for basic dynamical properties, and study the distribution function under the assumption of velocity isotropy. We find that none of the BPL models with realistic parameters has an isotropic distribution function that is positive over the entire phase space, implying that the BPL models cannot be supported by an isotropic velocity distribution, or models with a more radially anisotropic orbital structure. This result limits the attractiveness of the BPL family as a tool for lensing studies to some degree. More generally, we find that not all members of the general family of double power-law or Zhao models, often used to model dark matter haloes, can be supported by an isotropic or radially anisotropic distribution function. In other words, the distribution function may become negative even for spherically symmetric models with a well-behaved density profile.
\end{abstract}


\section{Introduction}
 
Since the first observational indications almost a century ago \citep{1932BAN.....6..249O, 1933AcHPh...6..110Z}, and especially since the first detailed analyses of galaxy rotation curves \citep{1970ApJ...160..811F, 1978ApJ...225L.107R}, it has been realised that dark matter plays a central role in the evolution of galaxies and the Universe as a whole. It is believed that about 85\% of all the matter in the Universe consists of dark matter \citep{2016A&A...594A..13P}. 

On galaxy scales, dark matter is traditionally traced through its gravitational effect on the kinematics of a tracer population such as the cold gas disk in spiral galaxies. The characteristics of dark matter haloes can in principle be inferred from a detailed analysis of the observed gas kinematics \citep[e.g.,][]{2001ApJ...552L..23D, 2008AJ....136.2648D, 2003ApJ...583..732S, 2004MNRAS.351..903G, 2017A&A...605A..55A}. In practice, one usually assumes a parameterised dark matter density profile, and the parameters of this distribution are determined by fitting the rotation curve. Most studies adopt either models in which the density has a power-law behaviour at both small and large radii \citep{1997ApJ...490..493N, 1999MNRAS.310.1147M, 2014MNRAS.441.2986D, 2014MNRAS.443.3712H, 2015ApJ...800...15H}, or models with a power-law logarithmic density slope \citep{2004MNRAS.349.1039N, 2010MNRAS.402...21N, 2005ApJ...624L..85M, 2006AJ....132.2685M, 2006ApJ...645.1001P}. Both classes of models have been studied in detail, and their photometric and dynamical properties are often simple enough to be expressed analytically \citep{1996MNRAS.278..488Z, 2005MNRAS.358.1325C, 2005MNRAS.360..492E, 2009ApJ...690.1280V, 2012A&A...540A..70R, 2020MNRAS.499.2912F}. 

An alternative and powerful way to constrain the distribution of dark matter on galaxy scales is gravitational lensing \citep[for reviews, see][]{2006glsw.conf...91K, 2010ARA&A..48...87T}. Despite a number of degeneracies that cannot be avoided \citep{1985ApJ...289L...1F, 2014A&A...564A.103S}, the multiplicity, position, shape, and flux of the lensed images offer interesting constraints to infer the characteristics of both the foreground lens and the lensed background source \citep{2006ApJ...649..599K, 2010ApJ...724..511A, 2012ApJ...747L..15G, 2017MNRAS.464.4823B, 2018MNRAS.478.4738N}. The mass models used in gravitational lensing studies are often based on a simple prescription of the mass surface density profile, with the most popular ones the singular isothermal lens, the softened isothermal sphere, and the power-law model. The advantage of these models is that many of the most important lensing characteristics, such as the shear, deflection angle and magnification, can be calculated analytically \citep{1993ApJ...417..450K, 2012MNRAS.426..868S, 2014MNRAS.437.1051W, 2015A&A...580A..79T, 2019MNRAS.487.5143O}. It has been argued, however, that simple models such as a single power-law model lack the flexibility required for high-precision galaxy lensing modelling, in particular for time-delay studies aimed at a measurement of the Hubble constant \citep{2016MNRAS.456..739X, 2018MNRAS.474.4648S, 2020MNRAS.493.1725K}. Ideally, one would use lens mass models similar to those used in dynamical studies. The lensing properties of general double-power-law models or Einasto models are, however, complex and less ideal for practical applications \citep{2000ApJ...534...34W, 2002ApJ...566..652L, 2012A&A...540A..70R, 2012A&A...546A..32R, 2020MNRAS.499.2912F}. Ideally, a lensing mass model should be used that has the characteristics of realistic dark matter distributions in galaxies, and that allows for an analytical calculation of the magnifications and deflection angles.

Very recently, a novel model with these characteristics was proposed by \citet{2020ApJ...892...62D}. As the name suggests, their BPL (broken power-law) model is characterised by a density profile that is a combination of two pure power-law profiles that meet at a certain break radius. Under the assumption of an elliptically symmetric surface mass distribution, the deflection angle and magnification can be evaluated analytically. \citet{2020ApJ...892...62D} conclude that this new analytical mass model is an efficient and realistic lensing mass model for galactic and cosmological applications. Interestingly, this model has a density profile with the basic characteristics of the most commonly used models used in dynamical analyses of dark matter haloes. In fact, the BPL model belongs to the general class of double power-law or Zhao models that is often used to model dark halo models in dynamical studies \citep[e.g.,][]{2013MNRAS.433.2314H, 2014MNRAS.443.3712H, 2015ApJ...800...15H, 2014MNRAS.441.2986D, 2017A&A...605A..55A, 2017MNRAS.468.1005D, 2020MNRAS.491.4523F, 2020MNRAS.499.2912F, 2020ApJ...904...45H}. 

These characteristics call for a more detailed study of the dynamical properties of the family of BPL models. One important aspect of the BPL models not covered in the analysis of \citet{2020ApJ...892...62D} is the phase space distribution function, the most fundamental quantity in dynamical studies \citep{1986PhR...133..217D, 2008gady.book.....B}. If one assumes the most simple orbital structure, i.e., an isotropic velocity distribution, every spherical density profile corresponds to a unique distribution function. It is, however, not guaranteed that this isotropic distribution function is positive over the entire phase space. It remains to be seen whether the BPL models presented by \citet{2020ApJ...892...62D} are consistent with an isotropic distribution function. Our recent work on the dynamical properties of the S\'ersic and Nuker models \citep{2019A&A...626A.110B, 2020A&A...634A.109B} has demonstrated that the positivity of the distribution function is not always satisfied, even for apparently well-behaved models where this might be expected at first sight. These recent results inspire us for a more thorough investigation of the dynamical structure of the BPL models, and of the broader family of Zhao models in general. 

The remainder of this paper is structured as follows. In Sec.~{\ref{BPL.sec}} we introduce the BPL models, and we discuss their basic properties and dynamical structure. In Sec.~{\ref{Zhao.sec}} we expand the BPL to the more general family of Zhao models and we investigate how the dynamical structure of these model changes if we gradually decrease the smoothness parameter $\alpha$. Finally, in Sec.~{\ref{Discussion.sec}} we discuss the implications of our findings and summarise our results. 

\section{The BPL model}
\label{BPL.sec}

\subsection{Basic properties}

\begin{figure*}
\centering
\includegraphics[width=\textwidth]{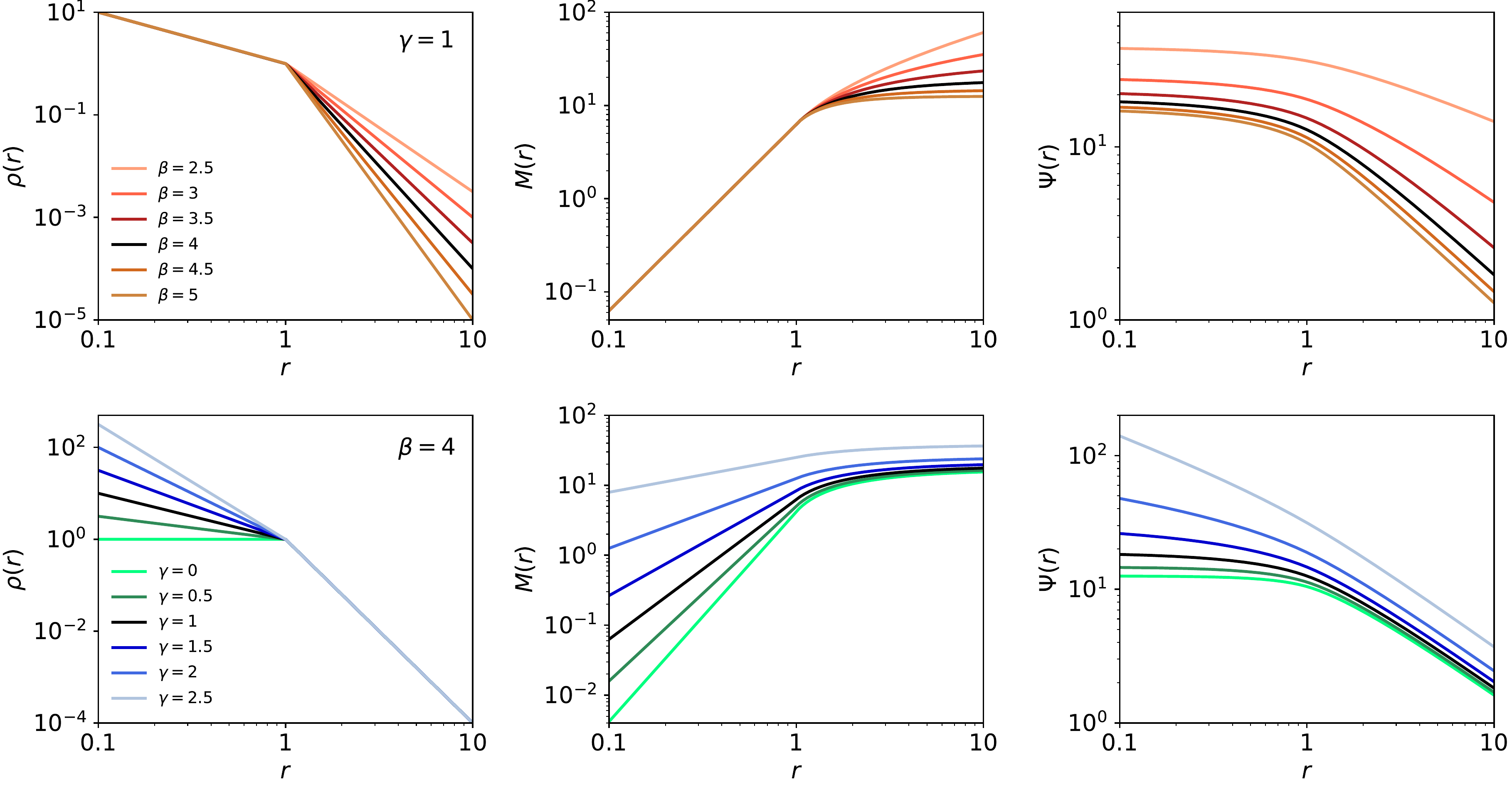}
\caption{Density profile $\rho(r)$, mass profile $M(r)$ and potential $\Psi(r)$ for different BPL models. The top panels shows models with $\gamma=1$ fixed and different values of $\beta$. The bottom row shows models with $\beta=4$ fixed and different values of $\gamma$. In all panels, the black curve corresponds to the model with $(\beta,\gamma) = (4,1)$.}
\label{BPL-density.fig}
\end{figure*}

The BPL model is characterised by the simple density profile
\begin{equation}
\rho(r) = \rho_{\text{b}}\times
\begin{cases}
\;(r/\rb)^{-\gamma} & \quad{\text{for }} r\leqslant\rb,\\
\;(r/\rb)^{-\beta} &\quad{\text{for }} r\geqslant\rb.
\end{cases}
\end{equation}
In this expression, $r$ is the spherical radius, $\rb$ is the break-radius that marks the transition between the inner and outer regions, $\rho_{\text{b}}$ is the density at the break radius, and $\beta$ and $\gamma$ are the negative logarithmic density slopes in the inner and outer regions, respectively. In their study, \citet{2020ApJ...892...62D} consider the range $0\leqslant \gamma<3$ for the inner density slope, and $1<\beta<3$ for the outer density slope. We consider the same range in $\gamma$, but consider the range $\beta>2$ for $\beta$. We do not consider models with $\beta\leqslant2$ as the gravitational potential for such models is ill-defined (see later). On the other hand, there is no compelling reason to limit the range of $\beta$ on the upper end. In fact, the range $\beta\geqslant3$ is probably the most interesting parameter range, as only models with $\beta>3$ have a finite total mass (see later), and the case $\beta=3$ corresponds to the most commonly used negative logarithmic density slope for dark matter haloes, such as the NFW model and the Moore profile \citep{1997ApJ...490..493N, 1999MNRAS.310.1147M}.

In the remainder of this paper we will work in dimensionless units with $G=\rb=\rho_{\text{b}}=1$. In these units, the density profile becomes
\begin{equation}
\rho(r) 
= 
\begin{cases}
\;r^{-\gamma} & \quad{\text{for }} r\leqslant 1,\\
\;r^{-\beta} & \quad{\text{for }} r\geqslant 1.
\end{cases}
\label{BPL-rho}
\end{equation}
The integrated mass profile $M(r)$ is readily found by integrating the density profile,
\begin{equation}
M(r) 
= 4\pi\int_0^r \rho(u)\,u^2\,\txd u.
\end{equation}
Inserting the expression~(\ref{BPL-rho}) for the density, we find
\begin{subequations}
\label{BPL-M}
\begin{equation}
M(r) 
=
\begin{cases}
\;M_{\text{in}}(r)
&\quad{\text{for }}r\leqslant 1,\\
\;M_{\text{out}}(r)
&\quad{\text{for }} r\geqslant 1,
\end{cases}
\end{equation}
with the inner mass profile given by
\begin{equation}
M_{\text{in}}(r) 
=
\frac{4\pi\,r^{3-\gamma}}{3-\gamma},
\end{equation}
and the outer profile by
\begin{equation}
M_{\text{out}}(r)
=
\begin{cases}
\;\dfrac{4\pi}{\beta-3}
\left(\dfrac{\beta-\gamma}{3-\gamma}-\dfrac{1}{r^{\beta-3}}\right)
&\quad{\text{if }}\beta\ne3,\\[1.2em]
\;4\pi\left(\dfrac{1}{3-\gamma} + \ln r\right)
&\quad{\text{if }} \beta=3.
\end{cases}
\end{equation}
\end{subequations}
The total mass is finite if $\beta>3$, and it that case, it is given by
\begin{equation}
M = 
\frac{4\pi\,(\beta-\gamma)}{(\beta-3)\,(3-\gamma)}.
\end{equation}
For a spherical mass distribution, the (positive) gravitational potential $\Psi(r)$ can be found through
\begin{equation}
\Psi(r) = 
\frac{M(r)}{r}+4\pi\int_r^\infty \rho(u)\,u\,\txd u.
\end{equation}
The latter integral diverges for all models with $\beta\leqslant2$, which is the reason why we limit our discussion to models with $\beta>2$. In analogy with the mass profile we write
\begin{subequations}
\label{BPL-Psi}
\begin{equation}
\Psi(r) 
=
\begin{cases}
\;\Psi_{\text{in}}(r)
&\quad{\text{for }}r\leqslant 1,\\
\;\Psi_{\text{out}}(r)
&\quad{\text{for }} r\geqslant 1.
\end{cases}
\end{equation}
with 
\begin{equation}
\Psi_{\text{in}}(r) 
= 
\begin{cases}
\;\dfrac{4\pi}{2-\gamma}\left(\dfrac{\beta-\gamma}{\beta-2}-\dfrac{r^{2-\gamma}}{3-\gamma}\right)
&\quad{\text{if }} \gamma\ne2,\\[1.2em]
\;4\pi\left(\dfrac{\beta-1}{\beta-2}-\ln r\right)
&\quad{\text{if }} \gamma=2,
\end{cases}
\end{equation}
and
\begin{equation}
\Psi_{\text{out}}(r) = 
\begin{cases}
\;\dfrac{4\pi}{\beta-3}\,\dfrac{1}{r}\left(\dfrac{\beta-\gamma}{3-\gamma}-\dfrac{r^{3-\beta}}{\beta-2}\right)
&\quad{\text{if }} \beta\ne3,\\[1.2em]
\;\dfrac{4\pi}{r}\left(\dfrac{4-\gamma}{3-\gamma}+\ln r\right)
&\quad{\text{if }} \beta=3.
\end{cases}
\end{equation}
\end{subequations}
The inner and outer profile merge smoothly at the break radius $r=1$, where the potential has the value
\begin{equation}
\Psib = \frac{4\pi\,(\beta-\gamma+1)}{(\beta-2)\,(3-\gamma)}.
\end{equation}
The potential well of the BPL models is infinitely deep for $\gamma\geqslant 2$, and finite for $\gamma<2$, with 
\begin{equation}
\Psi_0 \equiv \Psi(0) = \frac{4\pi\,(\beta-\gamma)}{(\beta-2)\,(2-\gamma)}.
\end{equation}
All models with $\beta>3$ show the expected Keplerian behaviour, $\Psi(r)\approx M/r$, at large radii. For the models with $\beta\leqslant3$, the total mass is infinite, and the gravitational potentials decreases less steeply than Keplerian at large radii.

Fig.~{\ref{BPL-density.fig}} shows the density profile, mass profile and gravitational potential for several BPL models with different inner and outer slopes. In the top panels, we vary the parameter $\beta$ while we keep $\gamma=1$ fixed, and in the bottom panels we keep $\beta=4$ and we vary $\gamma$. The black line in all panels corresponds to our fiducial model with $(\beta,\gamma)=(4,1)$, which is discussed in more detail in Appendix~{\ref{AppendixA.sec}}.

The dependence of the density, mass profile and potential immediately reflects the formulae~(\ref{BPL-rho}), (\ref{BPL-M}) and (\ref{BPL-Psi}). Note that, while the density profile shows a sharp break at $r=1$, the integrated mass profile and potential are smoother at the break radius. 

\subsection{Velocity dispersions}

\begin{figure*}
\centering
\includegraphics[width=\textwidth]{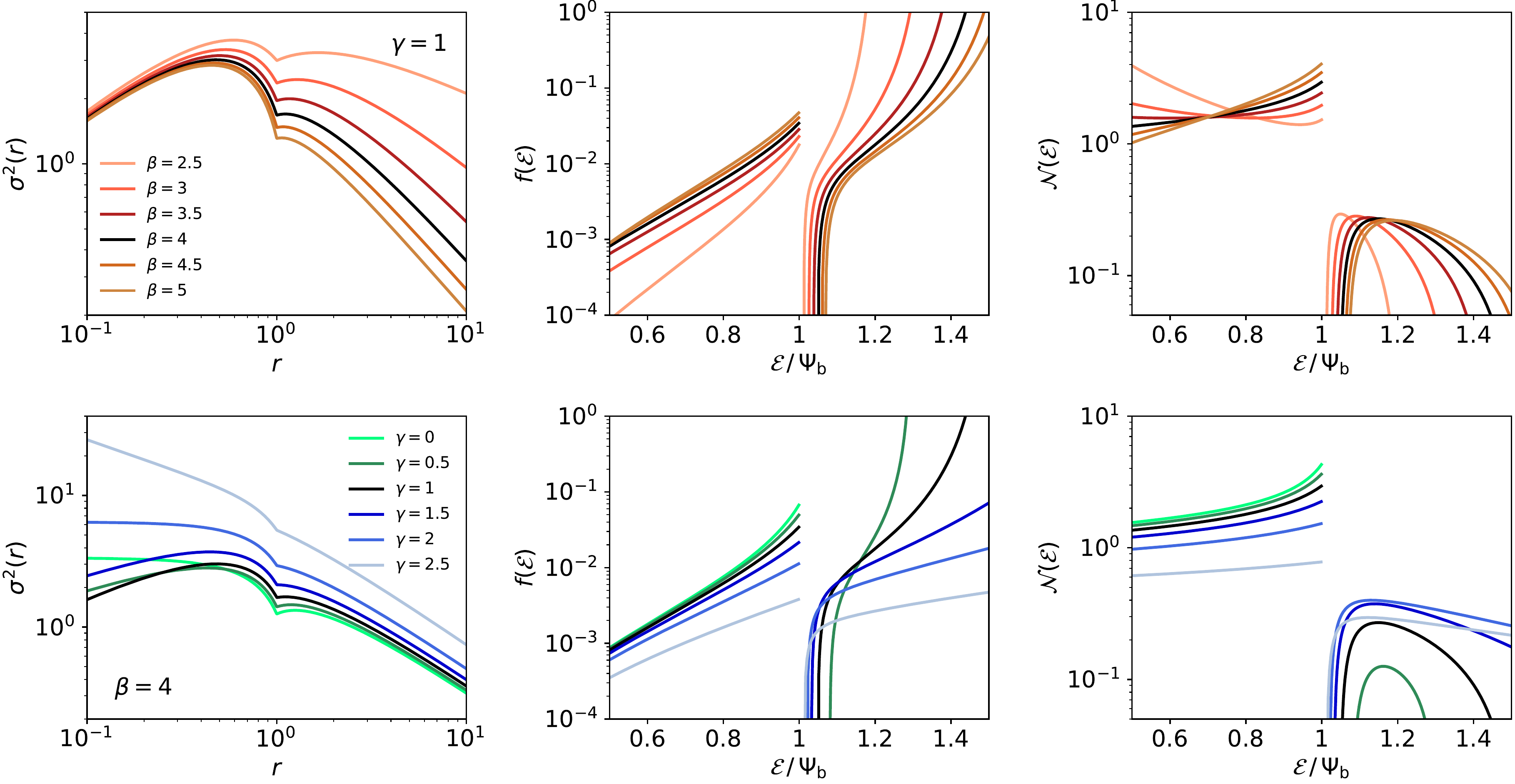}
\caption{Velocity dispersion profile $\sigma^2(r)$, phase-space distribution function $f(\calE)$ and differential energy distribution $\calN(\calE)$ for same BPL models as shown in Fig.~{\ref{BPL-density.fig}}. The color coding is the same as in the previous figure.}
\label{BPL-df.fig}
\end{figure*}

The main goal of this paper is to investigate whether the BPL model can be supported self-consistently by an isotropic velocity distribution. A first dynamical quantity to look at is the velocity dispersion profile. For isotropic dynamical models, the velocity dispersion profile $\sigma(r)$ can be found via the solution of the Jeans equation,
\begin{equation}
\sigma^2(r) = \frac{1}{\rho(r)} \int_r^\infty \frac{\rho(u)\,M(u)\,\txd u}{u^2}.
\label{Jeanssol}
\end{equation}
Using the expressions~(\ref{BPL-rho}) and (\ref{BPL-M}), we find the usual split between an inner and outer profile, 
\begin{subequations}
\label{BPL-sigma2}
\begin{equation}
\sigma^2(r) 
=
\begin{cases}
\;\sigma^2_{\text{in}}(r)
&\quad{\text{for }}r\leqslant 1,\\
\;\sigma^2_{\text{out}}(r)
&\quad{\text{for }} r\geqslant 1.
\end{cases}
\end{equation}
with
\begin{equation}
\sigma^2_{\text{in}}(r)
=
\begin{cases}
\;
\dfrac{2\pi\,r^\gamma}{(3-\gamma)\,(1-\gamma)}
\\[1em]
\quad\times
\left[\dfrac{(\beta-\gamma)\,(\beta-\gamma+2)}{\beta^2-1}-r^{2-\gamma}\right]
&\quad{\text{if }} \gamma\ne1,\\[1.2em]
\;
2\pi\,r\left(\dfrac{\beta}{\beta^2-1}-\ln r\right)
&\quad{\text{if }} \gamma=1.
\end{cases}
\end{equation}
and 
\begin{equation}
\sigma^2_{\text{out}}(r)
=
\begin{cases}
\;
\dfrac{2\pi}{(\beta-3)\,r}
\\[1em]
\quad\times
\left[\dfrac{2\,(\beta-\gamma)}{(\beta+1)\,(3-\gamma)}-\dfrac{r^{3-\beta}}{\beta-1}\right]
&\quad{\text{if }} \beta\ne3,\\[1.2em]
\;
\dfrac{\pi}{r}\left[\dfrac{7-\gamma}{4\,(3-\gamma)}+\ln r\right]
&\quad{\text{if }} \beta=3.
\end{cases}
\end{equation}
\end{subequations}
In the left panels of Fig.~{\ref{BPL-df.fig}} we show the dispersion profiles for the same models as discussed in Fig.~{\ref{BPL-density.fig}}. The dispersion profiles show an interesting variety in behaviour at the central regions for different values of $\gamma$. For $\gamma=0$, the dispersion converges to a finite value. For $0<\gamma<1$ the dispersion profile tends to zero as $\sigma^2\propto r^{\gamma}$, until it reaches an almost linear slope $\sigma^2\propto r\ln\tfrac1r$ for $\gamma=1$. For $1<\gamma<2$ the dispersion profile drops to zero as $\sigma^2\propto r^{2-\gamma}$, and it assumes a contant non-zero value again for $\gamma=2$. For values of $\gamma>2$ the dispersion grows infinitely large according to $\sigma^2\propto r^{-(\gamma-2)}$. In fact, this behaviour is not surprising, as it is exactly the same as for the $\gamma$-models \citep{1993MNRAS.265..250D, 1994AJ....107..634T}, and more generally, the entire family of Zhao models \citep{1996MNRAS.278..488Z}. Another conspicuous feature of the dispersion profiles is the very unnatural dip around the break radius, which obviously is the consequence of the sharp break in the density profile, and more specifically of the fact that the density appears in the denominator in expression~(\ref{Jeanssol}).

\subsection{Distribution function}

We now turn to the critical question whether the unique isotropic distribution function $f(\calE)$ is positive over the entire phase space. For a given density $\rho(r)$ and potential $\Psi(r)$, $f(\calE)$ can be found through Eddington's formula,
\begin{equation}
f(\calE) = \frac{1}{2\sqrt2\,\pi^2} \int_0^\calE \frac{D_\Psi^2\rho(\Psi)\,\txd\Psi}{\sqrt{\calE-\Psi}},
\label{Eddington}
\end{equation}
with $\rho(\Psi)$ the augmented density, i.e.\ the density written as a function of the potential. In principle, the potential (\ref{BPL-Psi}) of the BPL model can be inverted to a relation $r(\Psi)$ and combining this with the density (\ref{BPL-rho}) yields the augmented density. However, the inversion cannot be done analytically and the resulting formulae are cumbersome. A more elegant method is to follow \citet{1982MNRAS.200..951B} and \citet{2019A&A...626A.110B}, and recast expression~(\ref{Eddington}) as
\begin{equation}
f(\calE) = \frac{1}{2\sqrt2\,\pi^2} \int_{r_\calE}^\infty \frac{\Delta(r)\,\txd r}{\sqrt{\calE-\Psi(r)}},
\label{Eddington2}
\end{equation}
with 
\begin{equation}
\Delta(r) = \frac{r^2}{M(r)}\left[
D_r^2\rho(r) + D_r\rho(r)
\left(\frac{2}{r} - \frac{4\pi\,\rho(r)\,r^2}{M(r)}\right)
\right],
\label{defDelta}
\end{equation}
and $r_\calE$ the maximum radius that can be reached by a particle with binding energy $\calE$, which can be found through the implicit equation $\Psi(r_\calE) = \calE$. If we want to evaluate the distribution function, we first need to evaluate the first and second order derivative of the density profile. The derivative of the density profile~(\ref{BPL-rho}) is
\begin{equation}
D_r\rho(r) 
= 
\begin{cases}
\;-\gamma\,r^{-\gamma-1} &\quad {\text{for }} r< 1,\\
\;-\beta\,r^{-\beta-1} &\quad {\text{for }} r> 1.
\end{cases}
\label{BPL-drho}
\end{equation}
Note that the derivative is now a discontinuous function and hence not formally differentiable. However, we can write equation~(\ref{BPL-drho}) formally as 
\begin{equation}
D_r\rho(r) 
= 
-\gamma\,r^{-\gamma-1} 
- 
\left(\beta\,r^{-\beta-1} - \gamma\,r^{-\gamma-1}\right)
\Theta(r-1)
\label{BPL-drho2}
\end{equation}
with $\Theta(x)$ Heaviside's step function. As a consequence, the second derivative can be written as
\begin{multline}
D_r^2\rho(r) 
= 
-(\beta-\gamma)\,\delta(r-1)
\\
+
\begin{cases}
\;\gamma\,(\gamma+1)\,r^{-\gamma-2} &\quad {\text{for }} r< 1,\\
\;\beta\,(\beta+1)\,r^{-\beta-2} &\quad {\text{for }} r>1.
\end{cases}
\label{BPL-d2rho}
\end{multline}
When we combine the expressions~(\ref{Eddington2}), (\ref{defDelta}), (\ref{BPL-rho}), (\ref{BPL-M}), (\ref{BPL-drho2})  and (\ref{BPL-d2rho}), we find an expression for the distribution function. We write it as
\begin{subequations}
\label{BPL-df}
\begin{equation}
f(\calE) 
=
\begin{cases}
\;f_{\text{out}}(\calE)
&\quad{\text{for }}\calE<\Psib,\\
\;f_{\text{in}}(\calE)
&\quad{\text{for }} \Psib<\calE<\Psi_0.
\end{cases}
\end{equation}
where the outer profile now corresponds to binding energies below $\Psib$, and the inner profile to binding energies between $\Psib$ and the maximum value $\Psi_0$. For the outer profile we find
\begin{multline}
f_{\text{out}}(\calE) 
=
\dfrac{\beta\,(\beta-3)\,(3-\gamma)}{8\sqrt2\,\pi^3} 
\\
\times\int_{r_\calE}^\infty 
\dfrac{(\beta-1)\,(\beta-\gamma)\,u^\beta - 2\,(3-\gamma)\,u^3\,\txd u}{[(\beta-\gamma)\,u^\beta - (3-\gamma)\,u^3]^2 \sqrt{\calE-\Psi(u)}}. 
\label{BPL-df1}
\end{multline}
The expression for the inner distribution function is slightly more complicated,
\begin{align}
f_{\text{in}}(\calE)
=
&-\dfrac{(\beta-\gamma)\,(3-\gamma)}{8\sqrt{2}\,\pi^3}\,\frac{1}{\sqrt{\calE-\Psib}}
\nonumber \\
&+
\frac{(3-\gamma)\,\gamma}{4\sqrt2\,\pi^3} \int_{r_\calE}^1 \frac{\txd u}{u^3 \sqrt{\calE-\Psi(u)}} 
\label{BPL-dfinn}
\\[0.5em]
&+\dfrac{\beta\,(\beta-3)\,(3-\gamma)}{8\sqrt2\,\pi^3} 
\nonumber \\
&\qquad\times 
\int_1^\infty 
\dfrac{(\beta-1)\,(\beta-\gamma)\,u^\beta - 2\,(3-\gamma)\,u^3\,\txd u}{[(\beta-\gamma)\,u^\beta - (3-\gamma)\,u^3]^2 \sqrt{\calE-\Psi(u)}} .
\nonumber
\end{align}
\end{subequations}
The distribution functions for our standard set of BPL models are shown in the panels on the central column of Fig.~{\ref{BPL-df.fig}}. For any combination of the parameters $\beta$ and $\gamma$ with $\beta>\gamma$, the distribution function is a discontinuous function, with a prominent discontinuity at $\calE=\Psib$. Even more disturbing, however, is that $f(\calE)$ is negative at binding energies $\calE\gtrsim\Psib$. The first term in the expression for $\calE>\Psib$ is the culprit, as it shows an inverse square-root divergence to $-\infty$ as $\calE$ approaches $\Psib$ from the high binding energy side.

Only for the models with $2<\beta\leqslant\gamma<3$, the distribution function is positive over the entire range of binding energies. Models with $\beta=\gamma$ are pure power-law models with polytropic potential-density pairs. Their distribution function can be written as a pure power-law over the entire range of binding energy,
\begin{equation}
f(\calE) \propto \calE^{(6-\gamma)/(2\gamma-4)}.
\end{equation}
In the case $\beta<\gamma$, the distribution function is still discontinuous, but it is now positive over the entire phase space. The first term in expression~(\ref{BPL-dfinn}) now has a positive sign, implying that the distribution function diverges to $+\infty$ as $\calE$ approaches $\Psib$ from the high binding energy side. We note, however, that the models with $2<\beta\leqslant\gamma<3$, while academically interesting, are not realistic models for dark matter haloes. Not only do they necessarily have an unrealistically steep inner slope and an infinite total mass, they are also characterised by a steeper density slope in the central region compared to the outer region, which is not observed in simulated dark matter haloes. For the remainder of this paper, we assume that $\gamma<\beta$.

\subsection{Differential energy distribution}

The right panels of Fig.~{\ref{BPL-df.fig}} show the differential energy distribution $\calN(\calE)$ for the same BPL models as in the other panels. The differential energy distribution represents the distribution of mass as a function of the binding energy $\calE$. For isotropic systems, the differential energy distribution can be written as
\begin{equation}
\calN(\calE) = f(\calE)\,g(\calE),
\end{equation} 
with $g(\calE)$ the density-of-states function, defined as the phase-space volume accessible for a particle with binding energy $\calE$ \citep{2008gady.book.....B}. It can be calculated as
\begin{equation}
g(\calE) 
= 
16\sqrt2 \,\pi^2 \int_\calE^{\Psi_0} 
\left| r^2\,D_\Psi r\right| \sqrt{\Psi-\calE}\,\txd\Psi,
\label{gE}
\end{equation}
or equivalently
\begin{equation}
g(\calE) 
= 
16\sqrt2 \,\pi^2 \int_0^{r_\calE} 
r^2 \sqrt{\Psi(r)-\calE}\,\txd r.
\label{gE2}
\end{equation}
The behaviour of the differential energy distribution shows the same disturbing feature as the distribution function: it is positive over almost the entire range of binding energies, except for $\calE\gtrsim\Psib$ where it is negative for any value of $\beta$ or $\gamma$ (assuming $\gamma<\beta$). This again underlines the fact that the BPL model cannot be supported by an isotropic distribution function. 

The formal expression for the differential energy distribution is useful as a check for the calculations we have performed: given that $\calN(\calE)$ represents the distribution of mass as a function of the binding energy, it should satisfy the normalisation
\begin{equation}
\int_0^{\Psi_0} \calN(\calE)\,\txd\calE = M.
\end{equation}
A numerical integration of the differential energy distribution shown in the right panels of Fig.~{\ref{BPL-df.fig}} shows that this condition is indeed satisfied for all BPL models with a finite mass. 

\section{The Zhao models}
\label{Zhao.sec}

\subsection{Properties of the Zhao models}

\begin{figure*}
\centering
\includegraphics[width=\textwidth]{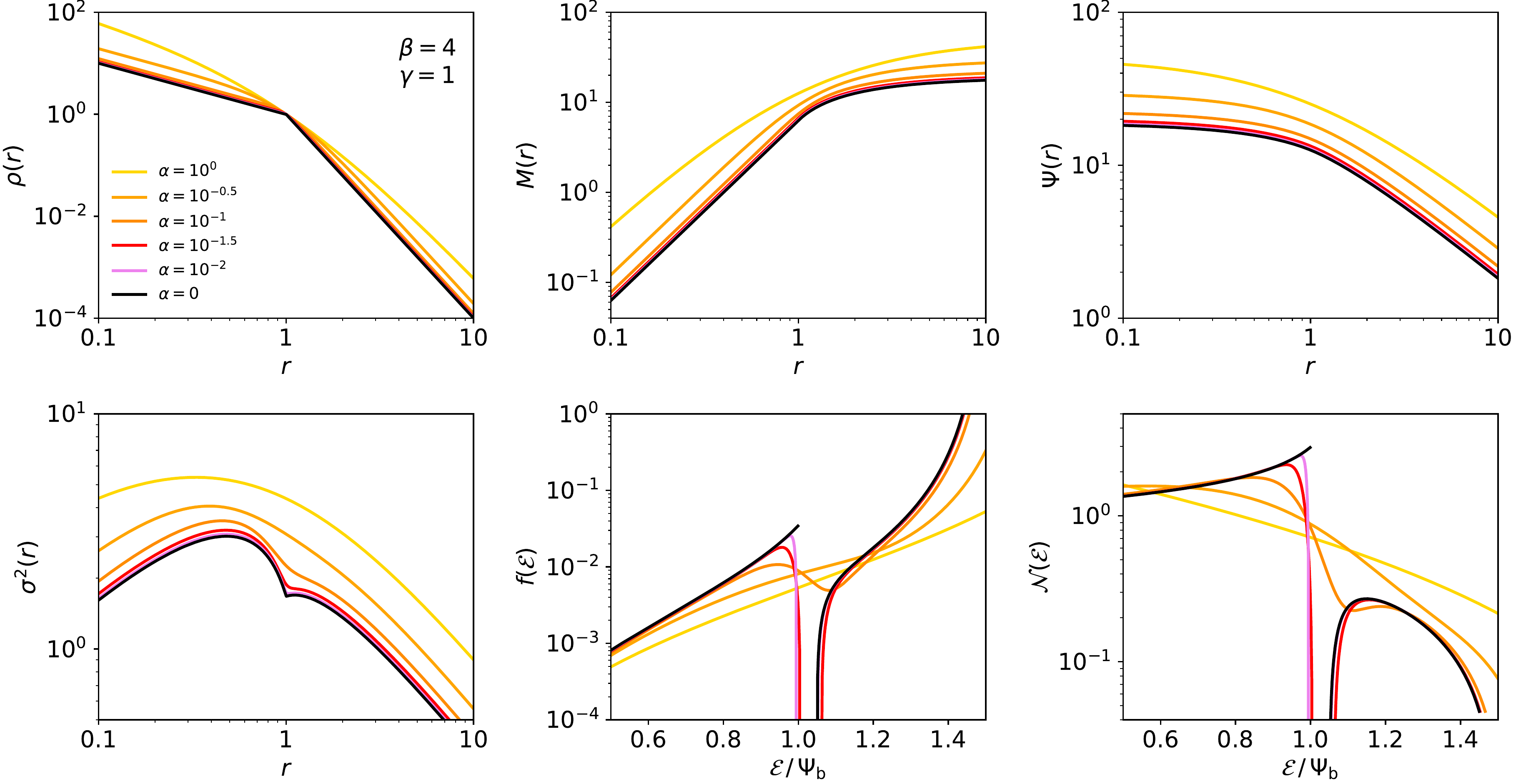}
\caption{Density, mass profile, potential, velocity dispersion, distribution function and differential energy distribution for different members of the subfamily of Zhao models with $(\beta,\gamma) = (4,1)$. The different models correspond to different values of the smoothness parameter $\alpha$ ranging between $\alpha=1$ and $\alpha=0$, corresponding to the BPL model with the same values of $\beta$ and $\gamma$.}
\label{Zhao41.fig}
\end{figure*}

The inconsistency of the isotropic BPL models raises the question whether the Zhao models can be  supported by an isotropic distribution function. The density profile of the Zhao model reads
\begin{equation}
\rho(r) = r^{-\gamma}\,\left(1+r^{1/\alpha}\right)^{-(\beta-\gamma)\,\alpha}.
\label{Zhao-rho}
\end{equation}
where we have adopted the same dimensionless units defined by $G = \rb = \rho_{\text{b}}=1$. The parameters $\beta$ and $\gamma$ are now the negative logarithmic density slopes in the limits $r\ll1$ and $r\gg1$, respectively. The third parameter, $\alpha$ corresponds to the smoothness of the transition between the inner and the outer regime. For large values of $\alpha$ this transition is very smooth and gradual, for decreasing values of $\alpha$ the transition becomes increasingly sharp. In the limit $\alpha=0$ the transition between the inner and outer profile becomes infinitely sharp and the density profile (\ref{Zhao-rho}) reduces to a broken power law, i.e. the density profile (\ref{BPL-rho}) of the BPL model.
The Zhao or $(\alpha,\beta,\gamma)$--models are very flexible and contain many different popular models as special cases, including the Plummer model \citep{1911MNRAS..71..460P, 1987MNRAS.224...13D}, the perfect sphere \citep{1985MNRAS.216..273D}, the modified Hubble sphere \citep{1972ApJ...175..627R, 2008gady.book.....B}, the Jaffe model \citep{1983MNRAS.202..995J}, the Hernquist model \citep{1990ApJ...356..359H, 2002A&A...393..485B}, the NFW model \citep{1997ApJ...490..493N}, and the Moore profile \citep{1999MNRAS.310.1147M}. A number of families of models, including the $\gamma$--models \citep{1993MNRAS.265..250D, 1994AJ....107..634T}, the Veltmann-models \citep{1979AZh....56..976V} and the Dekel-Zhao models \citep{2017MNRAS.468.1005D, 2020MNRAS.499.2912F} are subfamilies of this broad family.

\citet{1996MNRAS.278..488Z} presented several properties of this general family of models, focusing on the properties that can be calculated analytically. Explicit expressions for the isotropic distribution function are provided for a few cases, but no general discussion is offered. Interestingly, section~3.1 of the paper mentions that ``it can be easily shown that all of the $(\alpha,\beta,\gamma)$--models have a positive definite distribution function $f(\calE)\geqslant 0$ based on Eddington's formula.'' The fact that the isotropic distribution function for {\em{all}} BPL models is negative in some part of phase space casts doubt on that statement. 

\begin{figure*}
\centering
\includegraphics[width=0.75\textwidth]{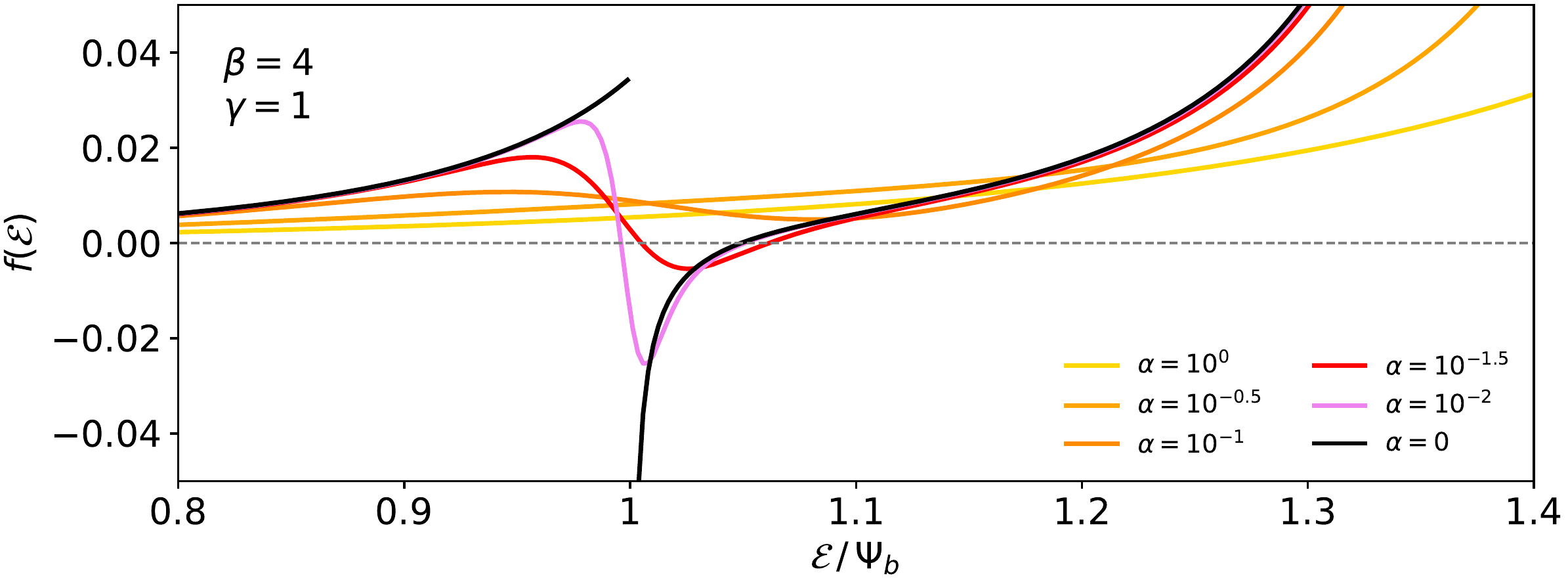}\hspace*{3em}
\\
\includegraphics[width=0.75\textwidth]{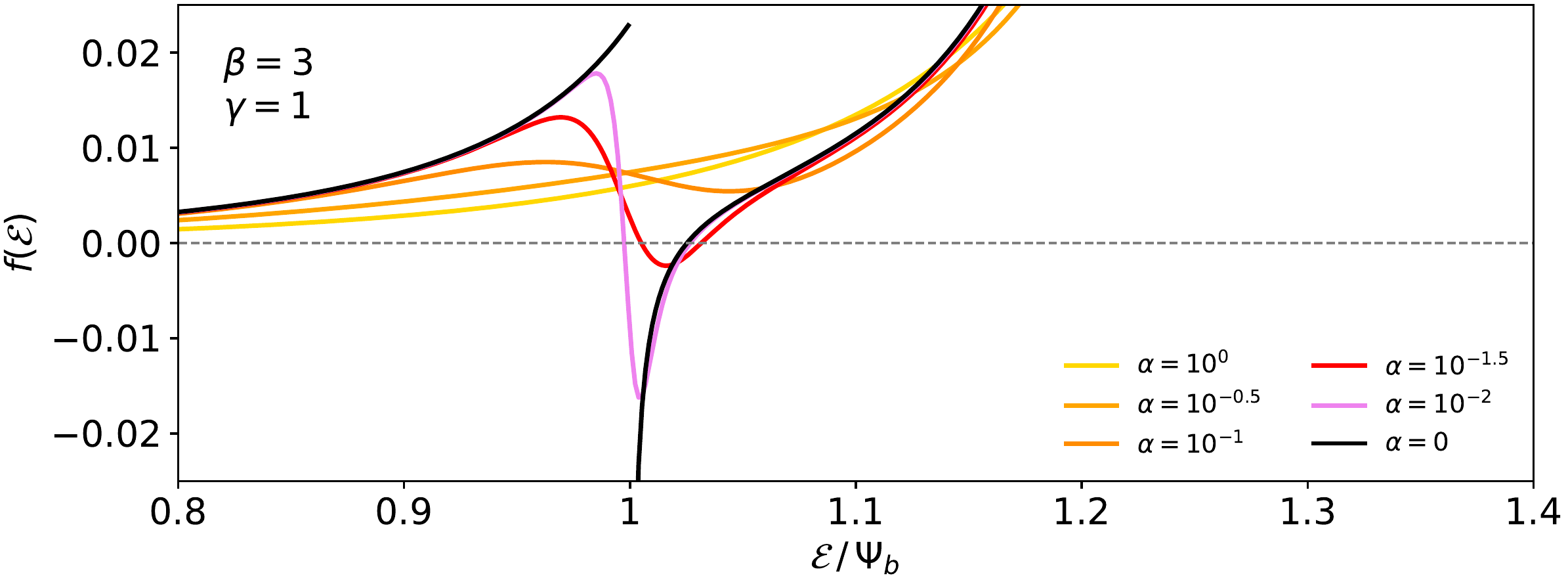}\hspace*{3em}
\\
\includegraphics[width=0.75\textwidth]{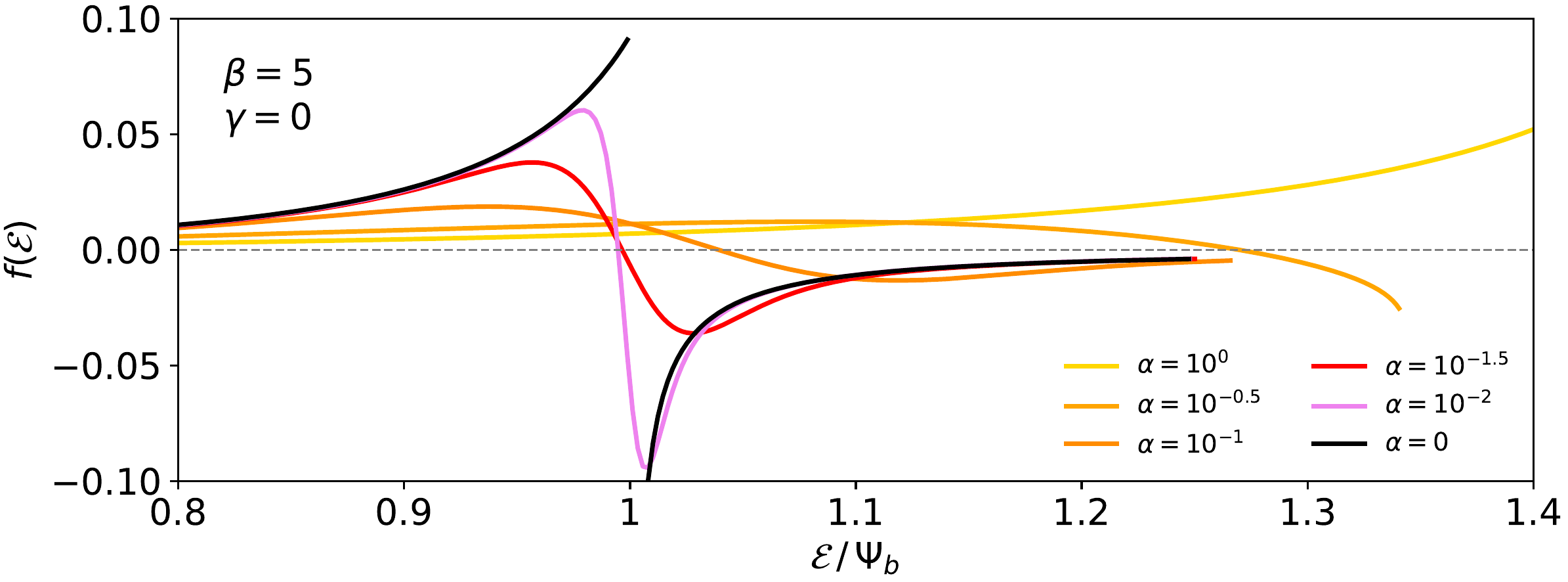}\hspace*{3em}
\caption{{\em{Top panel}}: Isotropic distribution functions $f(\calE)$ of generalised Hernquist models, corresponding to $(\beta,\gamma) = (4,1)$, for values of $\calE$ around $\calE=\Psib$. The distribution is plotted in linear scale to explicitly show potential negative values. {\em{Central panel}}: Same for generalised NFW models, corresponding to $(\beta,\gamma) = (3,1)$. {\em{Bottom panel}}: Same for generalised Plummer models, corresponding to $(\beta,\gamma) = (5,0)$.}
\label{Zhao41-dfzoom.fig}
\end{figure*}

\subsection{Evolution as a function of $\alpha$}

In Fig.~{\ref{Zhao41.fig}} we show the most important dynamical properties for a set of Zhao models with $(\beta,\gamma)=(4,1)$ and with different values of $\alpha$. All of these models have a mild $r^{-1}$ cusp and a $r^{-4}$ slope at large radii. As the Hernquist model is a special case of the general family of Zhao models corresponding to $(\alpha,\beta,\gamma) = (1,4,1)$, this subfamily can be regarded as a generalisation of the Hernquist model. In the limit $\alpha=0$ we recover our fiducial BPL model with $(\beta,\gamma) = (4,1)$.

To calculate the dynamical properties, we used a numerical Gauss-Legendre integrator with 128 nodes, implemented in C++. We tested our integration rules against high-precision numerical integration in Mathematica, and against analytical results available for a number of quantities and selected models. 

As $\alpha$ decreases, the density, mass profile and potential of the Zhao models shown nicely converge to the curves of the the corresponding BPL model. The same is true for the velocity dispersion profile, which gradually develops the conspicuous dip at $r=1$. Also for the distribution function and the differential energy distribution there is a clear convergence to the BPL curves. For large values of $\alpha$, such as for the Hernquist model with $\alpha=1$, the distribution function is a monotonically increasing function of $\calE$. As $\alpha$ decreases, the distribution function starts to develop a kink, corresponding to an excess at $\calE\lesssim\Psib$ and a depression at $\calE\gtrsim\Psib$. If $\alpha$ decreases even more, this kink becomes stronger and stronger, and at some point the distribution function drops below zero for $\calE\gtrsim\Psib$ . In the limit $\alpha=0$ we recover the result discussed in the previous section, with a distribution function that is discontinuous at $\calE=\Psib$, and that diverges to minus infinity when $\calE$ approaches $\Psib$ from the large binding energy side. 

This behaviour is more clearly illustrated in the top panel of Fig.~{\ref{Zhao41-dfzoom.fig}}, which adopts a linear rather than a logarithmic scale to allow for negative values to be shown. The models corresponding to the highest values of $\alpha$ (among which the Hernquist itself, corresponding to $\alpha=1$) have a distribution function that is a continuously increasing function of $\calE$. This implies not only that the isotropic model is self-consistent and physical, but also that it is stable against radial and non-radial perturbations \citep{1962spss.book.....A, 1971PhRvL..26..725D, 2008gady.book.....B}. The model with $\alpha=\tfrac{1}{10}$ has a distribution function with a kink and negative derivative at $\calE=\Psib$, but that still remains positive for all values of the binding energy. This means that the model is still physical, but that is no longer guaranteed to be stable against perturbations. On the other hand, the models with smallest values of $\alpha$ drop below zero. In these cases, the model cannot be supported by an isotropic distribution function. Finally, in the limit $\alpha=0$, the Zhao model reduces to the BPL model and the isotropic distribution function becomes discontinuous at $\calE=\Psib$ and negative for $\calE\gtrsim\Psib$.

To investigate whether this behaviour is a generic feature of the Zhao models and not connected to a particular choice of the parameters $\beta$ and $\gamma$, we have repeated this procedure for different values of $\beta$ and $\gamma$. The middle and the bottom panels of Fig.~{\ref{Zhao41-dfzoom.fig}} are similar to the top panel, but correspond to models with $(\beta,\gamma)=(3,1)$ and $(\beta,\gamma)=(5,0)$. These two subfamilies can be regarded as generalisations of the NFW and the Plummer model, respectively. Qualitatively speaking, the results are identical: for large $\alpha$ the distribution is a smooth, positive and monotonically increasing function of binding energy. As $\alpha$ gradually decreases, the distribution function starts to develop a kink. As soon as $\alpha$ drops below a critical value $\alpha_{\text{crit}}$, the exact value of which depends on $\beta$ and $\gamma$, the distribution function becomes negative for $\calE\gtrsim\Psib$. Finally, it turns into the discontinuous distribution function of the BPL model that diverges to $-\infty$ when $\calE$ approaches $\Psib$ from the high binding energy side. 

\begin{figure}
\includegraphics[width=0.92\columnwidth]{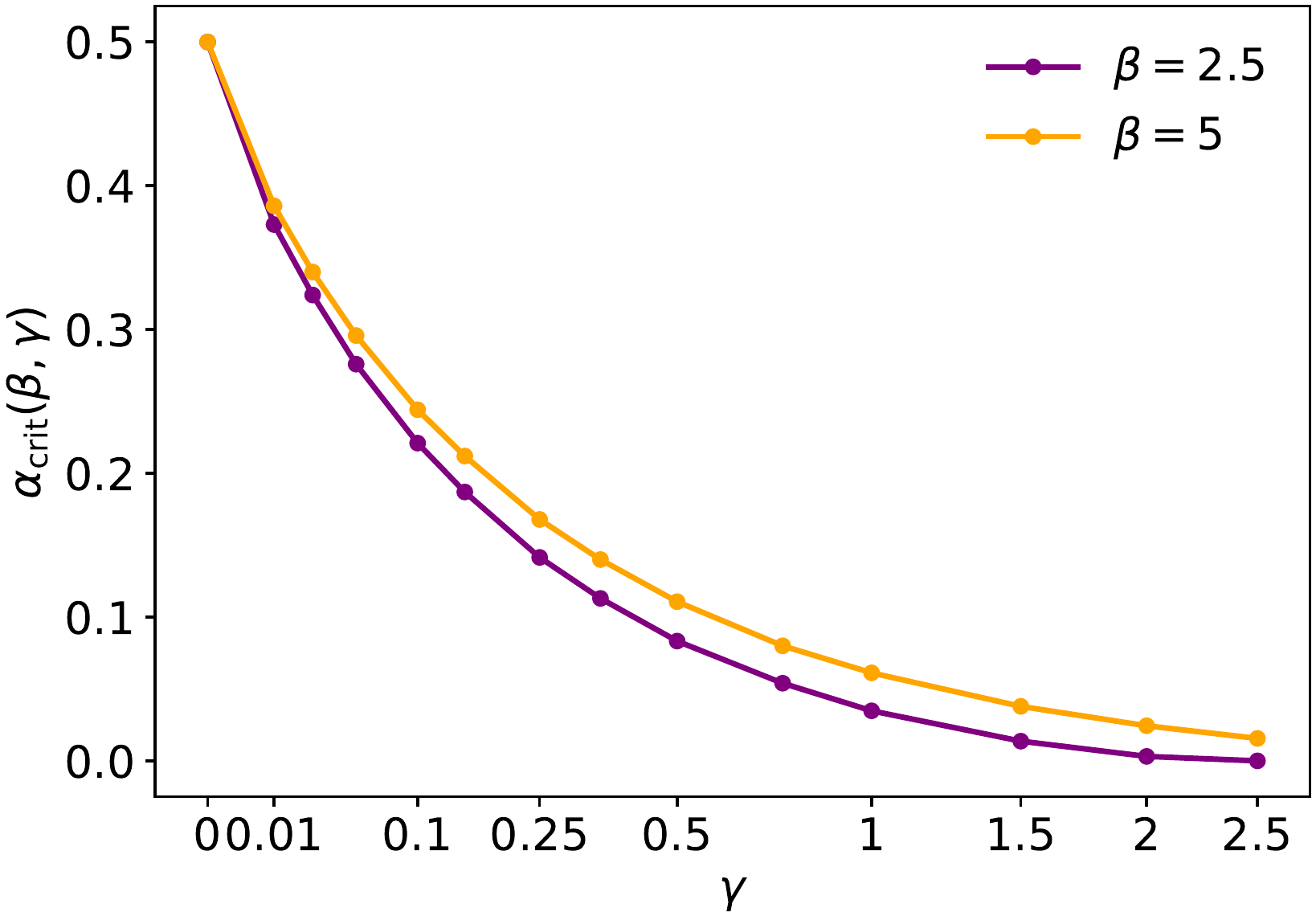}
\caption{$\alpha_{\text{crit}}(\beta,\gamma)$, the minimum value of $\alpha$ that still corresponds to a Zhao model with a positive distribution function $f(\calE)$, as a function of $\gamma$ for two different values of $\beta$.}
\label{Zhao-alphacrit.fig}
\end{figure}

In Fig.~{\ref{Zhao-alphacrit.fig}} we plot the $\alpha_{\text{crit}}(\beta,\gamma)$, the minimum value of $\alpha$ that still allows a positive definite isotropic distribution function, as a function of $\beta$ and $\gamma$, and in Table~{\ref{Zhao-alphacrit.tab}} we list a number of values for integer and half-integer values. For any value of $\beta$, $\alpha_{\text{crit}}(\beta,0) = \tfrac12$, that is, all models of the form
\begin{equation}
\rho(r) = (1+r^2)^{-\beta/2}
\end{equation}
are critical models in this respect. This includes the Plummer model for $\beta=5$, the perfect sphere for $\beta=4$, and the modified Hubble profile for $\beta=3$. $\alpha_{\text{crit}}$ quickly decreases as a function of $\gamma$, and already takes values around 0.1 for $\gamma=\tfrac12$. For any value of $\gamma$, $\alpha_{\text{crit}}$ is a weak and monotonically increasing function of $\beta$. All of the commonly used special cases of the general Zhao family with $\gamma>0$, such as the $\gamma$-models, the Dekel-Zhao models, and the Veltmann models have $\alpha>\alpha_{\text{crit}}(\beta,\gamma)$ and hence correspond to positive distribution functions $f(\calE)$.

However, our analysis clearly disproves the statement by \citet{1996MNRAS.278..488Z} that all $(\alpha,\beta,\gamma)$ models have a positive definite isotropic distribution function. In a follow-up paper, \citet{1997MNRAS.287..525Z} presents analytical fitting functions for the distribution function of his family of models. Actually, he proposes a fitting function for the primitive of the distribution function,
\begin{equation}
G(\calE) \equiv \int_0^\calE f(\calE')\,\txd\calE'.
\end{equation}
Inspired by the observation that the distribution function of asymptotic power-law systems is often, but not always, a power law of binding energy at asymptotic small or large radii, he proposes for $G(\calE)$ a similar double power-law function as for the density profile,
\begin{equation}
G(\calE) \approx f_0\,\Psib\,q^{\beta_2}\left(1+q^{1/\alpha_2}\right)^{(\gamma_2-\beta_2)\alpha_2}.
\end{equation}
In this expression, $q$ is a rescaled and dimensionless version of the binding energy,
\begin{equation}
q = 
\begin{cases}
\;\dfrac{\calE/\Psib}{1-\calE/\Psi_0} & \quad{\text{if }}\gamma<2, \\[0.75em]
\;\calE/\Psib & \quad{\text{if }}\gamma\geqslant2,
\end{cases}
\end{equation}
and $f_0$, $\alpha_2$, $\beta_2$ and $\gamma_2$ are four fitting parameters. By construction, this proposed fitting function for the distribution function is always positive definite and monotonically increasing as a function of binding energy. \citet{1997MNRAS.287..525Z} discusses a number of models where the proposed density derived by integrating this model distribution over velocity space describes the actual density profile very well with small residuals. All of these models correspond to safe values of $\alpha>\alpha_{\text{crit}}$ where the actual distribution is indeed positive and monotonically increasing. For smaller values of $\alpha$, for which the distribution function can show a kink or even assume negative values (see Fig.~{\ref{Zhao41-dfzoom.fig}}), the proposed fitting function cannot adequately describe the actual distribution function.

\begin{table}
\centering
\begin{tabular}{ccccccc}
\hline\\[-1em] 
$\beta$ & \multicolumn{6}{c}{ $\alpha_{\text{crit}}(\beta,\gamma)$} \\
& $\gamma=0$ & $\gamma=0.5$ & $\gamma=1$ & $\gamma=1.5$ 
& $\gamma=2$ & $\gamma=2.5$ \\[0.5em]
\hline \\[-0.5em] 
2.5 & 0.5000 & 0.0833 & 0.0348 & 0.0137 & 0.0031 & 0.0000 \\
3    & 0.5000 & 0.0919 & 0.0429 & 0.0207 & 0.0087 & 0.0016 \\
3.5 & 0.5000 & 0.0984 & 0.0490 & 0.0264 & 0.0136 & 0.0058 \\
4    & 0.5000 & 0.1034 & 0.0540 & 0.0309 & 0.0178 & 0.0095 \\ 
4.5 & 0.5000 & 0.1074 & 0.0579 & 0.0348 & 0.0213 & 0.0127 \\
5    & 0.5000 & 0.1107 & 0.0612 & 0.0379 & 0.0244 & 0.0156 \\[0.5em]
\hline
\end{tabular}
\caption{Numerical values of $\alpha_{\text{crit}}(\beta,\gamma)$, the minimum value of $\alpha$ that still corresponds to a Zhao model with a positive distribution function $f(\calE)$, for different values of $\beta$ and $\gamma$.}
\label{Zhao-alphacrit.tab}
\end{table}

\section{Discussion and conclusion}
\label{Discussion.sec}

The main goal of this paper was to further explore the set of BPL models introduced by \citet{2020ApJ...892...62D}. The prime specific question we wanted to address in this paper was whether the BPL models can be supported by an isotropic distribution function. Our analysis clearly answers negatively to this question:  {\em{none}} of the BPL models with realistic parameters (i.e., with $\gamma<\beta$) has a distribution function that is positive over the entire phase space. In all cases, the distribution function of the BPL model is a discontinuous function, with a discontinuity at $\calE=\Psib$, the binding energy corresponding to the value of the potential at the break radius. The distribution function is a well-behaved, smoothly increasing function of $\calE$ for all $\calE<\Psib$, but shows an inverse square-root divergence to $-\infty$ as $\calE$ approaches $\Psib$ from the high binding energy side. The BPL models can hence not be supported by an isotropic distribution function. 

More generally, we have found that not all Zhao models have positive definite isotropic distribution functions, contrary to the claim made by \citet{1996MNRAS.278..488Z}. Specifically, for every value of $\beta$ and $\gamma$ with $\gamma<\beta$, there is a minimum value $\alpha_{\text{crit}}(\beta,\gamma)$ that still allows positive distribution functions. Any model with $\alpha<\alpha_{\text{crit}}$, that is, with a sharper transition between the inner and outer regions, has a distribution function $f(\calE)$ that is negative at binding energies $\calE\gtrsim\Psib$. For $\gamma=0$, we find that $\alpha_{\text{crit}}(\beta,0)=\tfrac12$  for all values of $\beta$, implying that a number of well-known models such as the Plummer model, the perfect sphere and the modified Hubble sphere are critical models. For $\gamma>0$, all of the commonly used special cases of the general Zhao family are in the safe zone with $\alpha>\alpha_{\text{crit}}$. However, our analysis demonstrates that the positivity of the distribution function is a requirement that is not automatically satisfied, even for spherically symmetric models with a well-behaved density profile.

This fact is not completely unsurprising: in our previous work \citep{2019A&A...626A.110B, 2020A&A...634A.109B}, we have investigated the dynamical structure of the families of S\'ersic \citep{1968adga.book.....S, 1991A&A...249...99C} and Nuker \citep{2005AJ....129.2138L} models. Very similar to the Zhao models, both families are characterised by a different behaviour at small and large radii, and contain a shape parameter that governs the transition between these two asymptotic regimes (the S\'ersic parameter $m$ in the case of the S\'ersic models, and the smoothness parameter $\alpha$ in the case of the Nuker models). For both families we found that models with a smooth transition between the inner and outer regimes can be supported by an isotropic distribution function, but models with a sharp transition cannot. The family of Zhao models follows that general behaviour. The BPL models, as limiting cases of Zhao models with an infinitely sharp transition, automatically belong to the region in parameter space that corresponds to models that cannot be supported by an isotropic distribution function.

Why these models with a sharp transition in the density distribution cannot be supported by an isotropic distribution function can be understood by looking at a dynamical model as a superposition of orbits, that is, as in Schwartzschild orbit superposition methods \citep{1979ApJ...232..236S, 1984ApJ...286...27R, 2021MNRAS.500.1437N}. The weighted sum of all the orbits that pass a spherical shell at distance $r$ gives the density $\rho(r)$. In this thought experiment of building up a spherical model as a superposition of orbits, we start from the outer radius and gradually work towards the centre. Assume that we already have combination of orbits that reproduces the density and isotropic velocity structure at all radii $r$ beyond a given $\rref$. Obviously these orbits also cross into regions at small radii, and hence also contribute to the density at $r\leqslant \rref$. If we now want to ensure that our model has the correct density at $r=\rref$, we need to add orbits with apocentre at this radius (orbits with a smaller apocentre do not contribute to the density at $\rref$, and orbits with a larger apocentre would pollute the density at radii beyond $\rref$). To ensure velocity isotropy, we must add a mixture of such orbits to the set of orbits already present, with some of them more radial and some more circular-like. 

Repeating this for all radii, we can gradually build up a model that reproduces the density and the required isotropic velocity structure at all radii. There is one condition, however: at any radius $\rref$, the available ``room'' for orbits should not be filled yet previously. As long at the density increases sufficiently strongly as we move from large to small radii, there is always enough room to add orbits such that the density and the isotropic velocity structure is respected at all radii. Such models can be supported by an isotropic velocity distribution. However, in case there is a relatively sharp break in the density, say at $r=\rb$, there is less room to populate the model with orbits with apocentre at $r\sim \rb$. In case the break is very sharp, it is possible that the set of orbits that was necessary to guarantee the density and velocity isotropy at $r>\rb$ has already filled the entire space at $r\lesssim\rb$. The only way to ensure that the density and the velocity structure at these radii is respected consists of adding orbits with a negative weight to the mix. This results in a negative distribution function and differential energy distribution, which is obviously non-physical.  

Isotropy is only one possible orbital structure of dynamical systems, and there are many alternative options and methods to construct the corresponding distribution functions \citep[e.g.,][]{1979PAZh....5...77O, 1985AJ.....90.1027M, 1986PhR...133..217D, 1991MNRAS.253..414C}. It is possible to self-consistently generate the BPL density distributions with anisotropic distribution functions that favour tangential over radial orbits. In fact, any spherical density model can in theory be supported by a distribution function that consists of purely circular orbits \citep{1984ApJ...286...27R}. Cosmological N-body simulations generally yield dark matter haloes that are roughly isotropic in the inner regions and mildly radially anisotropic at larger regions \citep{2001ApJ...563..483T, 2004MNRAS.352..535D, 2011MNRAS.415.3895L, 2012ApJ...752..141L, 2013MNRAS.434.1576W, 2016MNRAS.462..663B}. If the BPL models cannot be supported by an isotropic velocity distribution, they cannot be supported by a radially anisotropic one either, as radial anisotropy puts tighter constraints on the positivity of the distribution function. This can be understood by looking at dynamical systems as a superposition of orbits, as discussed above. In a model with radial anisotropy, a majority of orbits are more elongated, and on average, each orbit with a given apocentre contributes more to the density at smaller radii. When gradually building up a model from outside inwards, at every radius $\rref$, a larger fraction of the available room for additional orbits will already be filled by orbits with apocentre $r>\rref$. A break in the density profile will therefore even more easily lead to the need to add orbits with negative weights. 

The family of BPL models was introduced as an attractive model for dark matter haloes in gravitational lensing studies. The attraction lies in two factors. First of all, in spite of its larger flexibility compared to more simple lens models, it still allows for most of the interesting properties for lensing studies to be calculated fully analytically, as demonstrated by \citet{2020ApJ...892...62D}. Secondly, contrary to most of the other popular analytical mass models used in lensing, the BPL model has the same global characteristics as the Zhao models, often used to model dark matter haloes \citep{1997ApJ...490..493N, 2013MNRAS.433.2314H, 2014MNRAS.443.3712H, 2015ApJ...800...15H, 2014MNRAS.441.2986D, 2015MNRAS.447.1353M, 2016MNRAS.456.3542T, 2017A&A...605A..55A, 2017MNRAS.468.1005D, 2017MNRAS.471..523W, 2017MNRAS.466.1648K, 2020MNRAS.494.4291C, 2020MNRAS.499.2912F, 2020ApJ...904...45H}. The fact that the BPL models cannot be supported by an isotropic (or radially anisotropic) distribution function limits the attractiveness of this family of models to some degree.

\section*{Data availability}

The data generated in this research will be shared on reasonable request to the corresponding author.

\section*{Acknowledgements}

We thank the referee, Dr Jonathan Freundlich, for a swift and constructive referee report that has led to an improved version of our work.

\bibliographystyle{mnras}
\bibliography{BPL}

\begin{thebibliography}{}
\makeatletter
\relax
\def\mn@urlcharsother{\let\do\@makeother \do\$\do\&\do\#\do\^\do\_\do\%\do\~}
\def\mn@doi{\begingroup\mn@urlcharsother \@ifnextchar [ {\mn@doi@}
  {\mn@doi@[]}}
\def\mn@doi@[#1]#2{\def\@tempa{#1}\ifx\@tempa\@empty \href
  {http://dx.doi.org/#2} {doi:#2}\else \href {http://dx.doi.org/#2} {#1}\fi
  \endgroup}
\def\mn@eprint#1#2{\mn@eprint@#1:#2::\@nil}
\def\mn@eprint@arXiv#1{\href {http://arxiv.org/abs/#1} {{\tt arXiv:#1}}}
\def\mn@eprint@dblp#1{\href {http://dblp.uni-trier.de/rec/bibtex/#1.xml}
  {dblp:#1}}
\def\mn@eprint@#1:#2:#3:#4\@nil{\def\@tempa {#1}\def\@tempb {#2}\def\@tempc
  {#3}\ifx \@tempc \@empty \let \@tempc \@tempb \let \@tempb \@tempa \fi \ifx
  \@tempb \@empty \def\@tempb {arXiv}\fi \@ifundefined
  {mn@eprint@\@tempb}{\@tempb:\@tempc}{\expandafter \expandafter \csname
  mn@eprint@\@tempb\endcsname \expandafter{\@tempc}}}

\bibitem[\protect\citeauthoryear{{Allaert}, {Gentile}  \& {Baes}}{{Allaert}
  et~al.}{2017}]{2017A&A...605A..55A}
{Allaert} F.,  {Gentile} G.,   {Baes} M.,  2017, \mn@doi [\aap]
  {10.1051/0004-6361/201730402}, \href
  {https://ui.adsabs.harvard.edu/abs/2017A&A...605A..55A} {605, A55}

\bibitem[\protect\citeauthoryear{{Antonov}}{{Antonov}}{1962}]{1962spss.book.....A}
{Antonov} V.~A.,  1962, {Solution of the problem of stability of stellar
  system}.
Vestnik Leningradskogo Universiteta, Leningrad

\bibitem[\protect\citeauthoryear{{Auger}, {Treu}, {Bolton}, {Gavazzi},
  {Koopmans}, {Marshall}, {Moustakas}  \& {Burles}}{{Auger}
  et~al.}{2010}]{2010ApJ...724..511A}
{Auger} M.~W.,  {Treu} T.,  {Bolton} A.~S.,  {Gavazzi} R.,  {Koopmans}
  L.~V.~E.,  {Marshall} P.~J.,  {Moustakas} L.~A.,   {Burles} S.,  2010,
  \mn@doi [\apj] {10.1088/0004-637X/724/1/511}, \href
  {https://ui.adsabs.harvard.edu/abs/2010ApJ...724..511A} {724, 511}

\bibitem[\protect\citeauthoryear{{Baes}}{{Baes}}{2020}]{2020A&A...634A.109B}
{Baes} M.,  2020, \mn@doi [\aap] {10.1051/0004-6361/201937209}, \href
  {https://ui.adsabs.harvard.edu/abs/2020A&A...634A.109B} {634, A109}

\bibitem[\protect\citeauthoryear{{Baes} \& {Ciotti}}{{Baes} \&
  {Ciotti}}{2019}]{2019A&A...626A.110B}
{Baes} M.,  {Ciotti} L.,  2019, \mn@doi [\aap] {10.1051/0004-6361/201935320},
  \href {https://ui.adsabs.harvard.edu/abs/2019A&A...626A.110B} {626, A110}

\bibitem[\protect\citeauthoryear{{Baes} \& {Dejonghe}}{{Baes} \&
  {Dejonghe}}{2002}]{2002A&A...393..485B}
{Baes} M.,  {Dejonghe} H.,  2002, \mn@doi [\aap] {10.1051/0004-6361:20021064},
  \href {https://ui.adsabs.harvard.edu/abs/2002A&A...393..485B} {393, 485}

\bibitem[\protect\citeauthoryear{{Baes} \& {van Hese}}{{Baes} \& {van
  Hese}}{2007}]{2007A&A...471..419B}
{Baes} M.,  {van Hese} E.,  2007, \mn@doi [\aap] {10.1051/0004-6361:20077672},
  \href {https://ui.adsabs.harvard.edu/abs/2007A&A...471..419B} {471, 419}

\bibitem[\protect\citeauthoryear{{Bellagamba}, {Tessore}  \&
  {Metcalf}}{{Bellagamba} et~al.}{2017}]{2017MNRAS.464.4823B}
{Bellagamba} F.,  {Tessore} N.,   {Metcalf} R.~B.,  2017, \mn@doi [\mnras]
  {10.1093/mnras/stw2726}, \href
  {https://ui.adsabs.harvard.edu/abs/2017MNRAS.464.4823B} {464, 4823}

\bibitem[\protect\citeauthoryear{{Binney}}{{Binney}}{1982}]{1982MNRAS.200..951B}
{Binney} J.,  1982, \mn@doi [\mnras] {10.1093/mnras/200.4.951}, \href
  {https://ui.adsabs.harvard.edu/abs/1982MNRAS.200..951B} {200, 951}

\bibitem[\protect\citeauthoryear{{Binney} \& {Tremaine}}{{Binney} \&
  {Tremaine}}{2008}]{2008gady.book.....B}
{Binney} J.,  {Tremaine} S.,  2008, {Galactic Dynamics: Second Edition}.
Princeton University Press

\bibitem[\protect\citeauthoryear{{Butsky} et~al.,}{{Butsky}
  et~al.}{2016}]{2016MNRAS.462..663B}
{Butsky} I.,  et~al., 2016, \mn@doi [\mnras] {10.1093/mnras/stw1688}, \href
  {https://ui.adsabs.harvard.edu/abs/2016MNRAS.462..663B} {462, 663}

\bibitem[\protect\citeauthoryear{{Cardone}, {Piedipalumbo}  \&
  {Tortora}}{{Cardone} et~al.}{2005}]{2005MNRAS.358.1325C}
{Cardone} V.~F.,  {Piedipalumbo} E.,   {Tortora} C.,  2005, \mn@doi [\mnras]
  {10.1111/j.1365-2966.2005.08834.x}, \href
  {http://adsabs.harvard.edu/abs/2005MNRAS.358.1325C} {358, 1325}

\bibitem[\protect\citeauthoryear{{Cautun} et~al.,}{{Cautun}
  et~al.}{2020}]{2020MNRAS.494.4291C}
{Cautun} M.,  et~al., 2020, \mn@doi [\mnras] {10.1093/mnras/staa1017}, \href
  {https://ui.adsabs.harvard.edu/abs/2020MNRAS.494.4291C} {494, 4291}

\bibitem[\protect\citeauthoryear{{Ciotti}}{{Ciotti}}{1991}]{1991A&A...249...99C}
{Ciotti} L.,  1991, \aap, \href
  {http://adsabs.harvard.edu/abs/1991A%26A...249...99C} {249, 99}

\bibitem[\protect\citeauthoryear{{Cuddeford}}{{Cuddeford}}{1991}]{1991MNRAS.253..414C}
{Cuddeford} P.,  1991, \mn@doi [\mnras] {10.1093/mnras/253.3.414}, \href
  {https://ui.adsabs.harvard.edu/abs/1991MNRAS.253..414C} {253, 414}

\bibitem[\protect\citeauthoryear{{Dehnen}}{{Dehnen}}{1993}]{1993MNRAS.265..250D}
{Dehnen} W.,  1993, \mn@doi [\mnras] {10.1093/mnras/265.1.250}, \href
  {http://adsabs.harvard.edu/abs/1993MNRAS.265..250D} {265, 250}

\bibitem[\protect\citeauthoryear{{Dejonghe}}{{Dejonghe}}{1986}]{1986PhR...133..217D}
{Dejonghe} H.,  1986, \mn@doi [\physrep] {10.1016/0370-1573(86)90098-0}, \href
  {https://ui.adsabs.harvard.edu/abs/1986PhR...133..217D} {133, 217}

\bibitem[\protect\citeauthoryear{{Dejonghe}}{{Dejonghe}}{1987}]{1987MNRAS.224...13D}
{Dejonghe} H.,  1987, \mn@doi [\mnras] {10.1093/mnras/224.1.13}, \href
  {http://adsabs.harvard.edu/abs/1987MNRAS.224...13D} {224, 13}

\bibitem[\protect\citeauthoryear{{Dekel}, {Ishai}, {Dutton}  \&
  {Maccio}}{{Dekel} et~al.}{2017}]{2017MNRAS.468.1005D}
{Dekel} A.,  {Ishai} G.,  {Dutton} A.~A.,   {Maccio} A.~V.,  2017, \mn@doi
  [\mnras] {10.1093/mnras/stx486}, \href
  {https://ui.adsabs.harvard.edu/abs/2017MNRAS.468.1005D} {468, 1005}

\bibitem[\protect\citeauthoryear{{Di Cintio}, {Brook}, {Dutton}, {Macci{\`o}},
  {Stinson}  \& {Knebe}}{{Di Cintio} et~al.}{2014}]{2014MNRAS.441.2986D}
{Di Cintio} A.,  {Brook} C.~B.,  {Dutton} A.~A.,  {Macci{\`o}} A.~V.,
  {Stinson} G.~S.,   {Knebe} A.,  2014, \mn@doi [\mnras]
  {10.1093/mnras/stu729}, \href
  {https://ui.adsabs.harvard.edu/abs/2014MNRAS.441.2986D} {441, 2986}

\bibitem[\protect\citeauthoryear{{Diemand}, {Moore}  \& {Stadel}}{{Diemand}
  et~al.}{2004}]{2004MNRAS.352..535D}
{Diemand} J.,  {Moore} B.,   {Stadel} J.,  2004, \mn@doi [\mnras]
  {10.1111/j.1365-2966.2004.07940.x}, \href
  {https://ui.adsabs.harvard.edu/abs/2004MNRAS.352..535D} {352, 535}

\bibitem[\protect\citeauthoryear{{Doremus}, {Feix}  \& {Baumann}}{{Doremus}
  et~al.}{1971}]{1971PhRvL..26..725D}
{Doremus} J.-P.,  {Feix} M.~R.,   {Baumann} G.,  1971, \mn@doi [Physical Review
  Letters] {10.1103/PhysRevLett.26.725}, \href
  {http://adsabs.harvard.edu/abs/1971PhRvL..26..725D} {26, 725}

\bibitem[\protect\citeauthoryear{{Du}, {Zhao}, {Fan}, {Shu}, {Li}  \&
  {Mao}}{{Du} et~al.}{2020}]{2020ApJ...892...62D}
{Du} W.,  {Zhao} G.-B.,  {Fan} Z.,  {Shu} Y.,  {Li} R.,   {Mao} S.,  2020,
  \mn@doi [\apj] {10.3847/1538-4357/ab7a15}, \href
  {https://ui.adsabs.harvard.edu/abs/2020ApJ...892...62D} {892, 62}

\bibitem[\protect\citeauthoryear{{Evans} \& {An}}{{Evans} \&
  {An}}{2005}]{2005MNRAS.360..492E}
{Evans} N.~W.,  {An} J.,  2005, \mn@doi [\mnras]
  {10.1111/j.1365-2966.2005.09078.x}, \href
  {http://adsabs.harvard.edu/abs/2005MNRAS.360..492E} {360, 492}

\bibitem[\protect\citeauthoryear{{Falco}, {Gorenstein}  \& {Shapiro}}{{Falco}
  et~al.}{1985}]{1985ApJ...289L...1F}
{Falco} E.~E.,  {Gorenstein} M.~V.,   {Shapiro} I.~I.,  1985, \mn@doi [\apjl]
  {10.1086/184422}, \href
  {https://ui.adsabs.harvard.edu/abs/1985ApJ...289L...1F} {289, L1}

\bibitem[\protect\citeauthoryear{{Freeman}}{{Freeman}}{1970}]{1970ApJ...160..811F}
{Freeman} K.~C.,  1970, \mn@doi [\apj] {10.1086/150474}, \href
  {https://ui.adsabs.harvard.edu/abs/1970ApJ...160..811F} {160, 811}

\bibitem[\protect\citeauthoryear{{Freundlich}, {Dekel}, {Jiang}, {Ishai},
  {Cornuault}, {Lapiner}, {Dutton}  \& {Macci{\`o}}}{{Freundlich}
  et~al.}{2020a}]{2020MNRAS.491.4523F}
{Freundlich} J.,  {Dekel} A.,  {Jiang} F.,  {Ishai} G.,  {Cornuault} N.,
  {Lapiner} S.,  {Dutton} A.~A.,   {Macci{\`o}} A.~V.,  2020a, \mn@doi [\mnras]
  {10.1093/mnras/stz3306}, \href
  {https://ui.adsabs.harvard.edu/abs/2020MNRAS.491.4523F} {491, 4523}

\bibitem[\protect\citeauthoryear{{Freundlich} et~al.,}{{Freundlich}
  et~al.}{2020b}]{2020MNRAS.499.2912F}
{Freundlich} J.,  et~al., 2020b, \mn@doi [\mnras] {10.1093/mnras/staa2790},
  \href {https://ui.adsabs.harvard.edu/abs/2020MNRAS.499.2912F} {499, 2912}

\bibitem[\protect\citeauthoryear{{Gentile}, {Salucci}, {Klein}, {Vergani}  \&
  {Kalberla}}{{Gentile} et~al.}{2004}]{2004MNRAS.351..903G}
{Gentile} G.,  {Salucci} P.,  {Klein} U.,  {Vergani} D.,   {Kalberla} P.,
  2004, \mn@doi [\mnras] {10.1111/j.1365-2966.2004.07836.x}, \href
  {https://ui.adsabs.harvard.edu/abs/2004MNRAS.351..903G} {351, 903}

\bibitem[\protect\citeauthoryear{{Grillo}}{{Grillo}}{2012}]{2012ApJ...747L..15G}
{Grillo} C.,  2012, \mn@doi [\apjl] {10.1088/2041-8205/747/1/L15}, \href
  {https://ui.adsabs.harvard.edu/abs/2012ApJ...747L..15G} {747, L15}

\bibitem[\protect\citeauthoryear{{Hague} \& {Wilkinson}}{{Hague} \&
  {Wilkinson}}{2013}]{2013MNRAS.433.2314H}
{Hague} P.~R.,  {Wilkinson} M.~I.,  2013, \mn@doi [\mnras]
  {10.1093/mnras/stt899}, \href
  {https://ui.adsabs.harvard.edu/abs/2013MNRAS.433.2314H} {433, 2314}

\bibitem[\protect\citeauthoryear{{Hague} \& {Wilkinson}}{{Hague} \&
  {Wilkinson}}{2014}]{2014MNRAS.443.3712H}
{Hague} P.~R.,  {Wilkinson} M.~I.,  2014, \mn@doi [\mnras]
  {10.1093/mnras/stu1376}, \href
  {https://ui.adsabs.harvard.edu/abs/2014MNRAS.443.3712H} {443, 3712}

\bibitem[\protect\citeauthoryear{{Hague} \& {Wilkinson}}{{Hague} \&
  {Wilkinson}}{2015}]{2015ApJ...800...15H}
{Hague} P.~R.,  {Wilkinson} M.~I.,  2015, \mn@doi [\apj]
  {10.1088/0004-637X/800/1/15}, \href
  {https://ui.adsabs.harvard.edu/abs/2015ApJ...800...15H} {800, 15}

\bibitem[\protect\citeauthoryear{{Hayashi}, {Chiba}  \& {Ishiyama}}{{Hayashi}
  et~al.}{2020}]{2020ApJ...904...45H}
{Hayashi} K.,  {Chiba} M.,   {Ishiyama} T.,  2020, \mn@doi [\apj]
  {10.3847/1538-4357/abbe0a}, \href
  {https://ui.adsabs.harvard.edu/abs/2020ApJ...904...45H} {904, 45}

\bibitem[\protect\citeauthoryear{{Hernquist}}{{Hernquist}}{1990}]{1990ApJ...356..359H}
{Hernquist} L.,  1990, \mn@doi [\apj] {10.1086/168845}, \href
  {http://adsabs.harvard.edu/abs/1990ApJ...356..359H} {356, 359}

\bibitem[\protect\citeauthoryear{{Jaffe}}{{Jaffe}}{1983}]{1983MNRAS.202..995J}
{Jaffe} W.,  1983, \mn@doi [\mnras] {10.1093/mnras/202.4.995}, \href
  {http://adsabs.harvard.edu/abs/1983MNRAS.202..995J} {202, 995}

\bibitem[\protect\citeauthoryear{{Kassiola} \& {Kovner}}{{Kassiola} \&
  {Kovner}}{1993}]{1993ApJ...417..450K}
{Kassiola} A.,  {Kovner} I.,  1993, \mn@doi [\apj] {10.1086/173325}, \href
  {https://ui.adsabs.harvard.edu/abs/1993ApJ...417..450K} {417, 450}

\bibitem[\protect\citeauthoryear{{Katz}, {Lelli}, {McGaugh}, {Di Cintio},
  {Brook}  \& {Schombert}}{{Katz} et~al.}{2017}]{2017MNRAS.466.1648K}
{Katz} H.,  {Lelli} F.,  {McGaugh} S.~S.,  {Di Cintio} A.,  {Brook} C.~B.,
  {Schombert} J.~M.,  2017, \mn@doi [\mnras] {10.1093/mnras/stw3101}, \href
  {https://ui.adsabs.harvard.edu/abs/2017MNRAS.466.1648K} {466, 1648}

\bibitem[\protect\citeauthoryear{{Kochanek}}{{Kochanek}}{2006}]{2006glsw.conf...91K}
{Kochanek} C.~S.,  2006, in {Meylan} G.,  {Jetzer} P.,  {North} P.,
  {Schneider} P.,  {Kochanek} C.~S.,   {Wambsganss} J.,  eds, Saas-Fee Advanced
  Course 33: Gravitational Lensing: Strong, Weak and Micro. pp 91--268

\bibitem[\protect\citeauthoryear{{Kochanek}}{{Kochanek}}{2020}]{2020MNRAS.493.1725K}
{Kochanek} C.~S.,  2020, \mn@doi [\mnras] {10.1093/mnras/staa344}, \href
  {https://ui.adsabs.harvard.edu/abs/2020MNRAS.493.1725K} {493, 1725}

\bibitem[\protect\citeauthoryear{{Koopmans}, {Treu}, {Bolton}, {Burles}  \&
  {Moustakas}}{{Koopmans} et~al.}{2006}]{2006ApJ...649..599K}
{Koopmans} L. V.~E.,  {Treu} T.,  {Bolton} A.~S.,  {Burles} S.,   {Moustakas}
  L.~A.,  2006, \mn@doi [\apj] {10.1086/505696}, \href
  {https://ui.adsabs.harvard.edu/abs/2006ApJ...649..599K} {649, 599}

\bibitem[\protect\citeauthoryear{{Lauer} et~al.,}{{Lauer}
  et~al.}{2005}]{2005AJ....129.2138L}
{Lauer} T.~R.,  et~al., 2005, \mn@doi [\aj] {10.1086/429565}, \href
  {https://ui.adsabs.harvard.edu/abs/2005AJ....129.2138L} {129, 2138}

\bibitem[\protect\citeauthoryear{{Lemze} et~al.,}{{Lemze}
  et~al.}{2012}]{2012ApJ...752..141L}
{Lemze} D.,  et~al., 2012, \mn@doi [\apj] {10.1088/0004-637X/752/2/141}, \href
  {https://ui.adsabs.harvard.edu/abs/2012ApJ...752..141L} {752, 141}

\bibitem[\protect\citeauthoryear{{Li} \& {Ostriker}}{{Li} \&
  {Ostriker}}{2002}]{2002ApJ...566..652L}
{Li} L.-X.,  {Ostriker} J.~P.,  2002, \mn@doi [\apj] {10.1086/338330}, \href
  {https://ui.adsabs.harvard.edu/abs/2002ApJ...566..652L} {566, 652}

\bibitem[\protect\citeauthoryear{{Ludlow}, {Navarro}, {White},
  {Boylan-Kolchin}, {Springel}, {Jenkins}  \& {Frenk}}{{Ludlow}
  et~al.}{2011}]{2011MNRAS.415.3895L}
{Ludlow} A.~D.,  {Navarro} J.~F.,  {White} S. D.~M.,  {Boylan-Kolchin} M.,
  {Springel} V.,  {Jenkins} A.,   {Frenk} C.~S.,  2011, \mn@doi [\mnras]
  {10.1111/j.1365-2966.2011.19008.x}, \href
  {https://ui.adsabs.harvard.edu/abs/2011MNRAS.415.3895L} {415, 3895}

\bibitem[\protect\citeauthoryear{{Merritt}}{{Merritt}}{1985}]{1985AJ.....90.1027M}
{Merritt} D.,  1985, \mn@doi [\aj] {10.1086/113810}, \href
  {https://ui.adsabs.harvard.edu/abs/1985AJ.....90.1027M} {90, 1027}

\bibitem[\protect\citeauthoryear{{Merritt}, {Navarro}, {Ludlow}  \&
  {Jenkins}}{{Merritt} et~al.}{2005}]{2005ApJ...624L..85M}
{Merritt} D.,  {Navarro} J.~F.,  {Ludlow} A.,   {Jenkins} A.,  2005, \mn@doi
  [\apjl] {10.1086/430636}, \href
  {https://ui.adsabs.harvard.edu/abs/2005ApJ...624L..85M} {624, L85}

\bibitem[\protect\citeauthoryear{{Merritt}, {Graham}, {Moore}, {Diemand}  \&
  {Terzi{\'c}}}{{Merritt} et~al.}{2006}]{2006AJ....132.2685M}
{Merritt} D.,  {Graham} A.~W.,  {Moore} B.,  {Diemand} J.,   {Terzi{\'c}} B.,
  2006, \mn@doi [\aj] {10.1086/508988}, \href
  {https://ui.adsabs.harvard.edu/abs/2006AJ....132.2685M} {132, 2685}

\bibitem[\protect\citeauthoryear{{Mollitor}, {Nezri}  \& {Teyssier}}{{Mollitor}
  et~al.}{2015}]{2015MNRAS.447.1353M}
{Mollitor} P.,  {Nezri} E.,   {Teyssier} R.,  2015, \mn@doi [\mnras]
  {10.1093/mnras/stu2466}, \href
  {https://ui.adsabs.harvard.edu/abs/2015MNRAS.447.1353M} {447, 1353}

\bibitem[\protect\citeauthoryear{{Moore}, {Quinn}, {Governato}, {Stadel}  \&
  {Lake}}{{Moore} et~al.}{1999}]{1999MNRAS.310.1147M}
{Moore} B.,  {Quinn} T.,  {Governato} F.,  {Stadel} J.,   {Lake} G.,  1999,
  \mn@doi [\mnras] {10.1046/j.1365-8711.1999.03039.x}, \href
  {https://ui.adsabs.harvard.edu/abs/1999MNRAS.310.1147M} {310, 1147}

\bibitem[\protect\citeauthoryear{{Navarro}, {Frenk}  \& {White}}{{Navarro}
  et~al.}{1997}]{1997ApJ...490..493N}
{Navarro} J.~F.,  {Frenk} C.~S.,   {White} S. D.~M.,  1997, \mn@doi [\apj]
  {10.1086/304888}, \href
  {https://ui.adsabs.harvard.edu/abs/1997ApJ...490..493N} {490, 493}

\bibitem[\protect\citeauthoryear{{Navarro} et~al.,}{{Navarro}
  et~al.}{2004}]{2004MNRAS.349.1039N}
{Navarro} J.~F.,  et~al., 2004, \mn@doi [\mnras]
  {10.1111/j.1365-2966.2004.07586.x}, \href
  {https://ui.adsabs.harvard.edu/abs/2004MNRAS.349.1039N} {349, 1039}

\bibitem[\protect\citeauthoryear{{Navarro} et~al.,}{{Navarro}
  et~al.}{2010}]{2010MNRAS.402...21N}
{Navarro} J.~F.,  et~al., 2010, \mn@doi [\mnras]
  {10.1111/j.1365-2966.2009.15878.x}, \href
  {https://ui.adsabs.harvard.edu/abs/2010MNRAS.402...21N} {402, 21}

\bibitem[\protect\citeauthoryear{{Neureiter} et~al.,}{{Neureiter}
  et~al.}{2021}]{2021MNRAS.500.1437N}
{Neureiter} B.,  et~al., 2021, \mn@doi [\mnras] {10.1093/mnras/staa3014}, \href
  {https://ui.adsabs.harvard.edu/abs/2021MNRAS.500.1437N} {500, 1437}

\bibitem[\protect\citeauthoryear{{Nightingale}, {Dye}  \&
  {Massey}}{{Nightingale} et~al.}{2018}]{2018MNRAS.478.4738N}
{Nightingale} J.~W.,  {Dye} S.,   {Massey} R.~J.,  2018, \mn@doi [\mnras]
  {10.1093/mnras/sty1264}, \href
  {https://ui.adsabs.harvard.edu/abs/2018MNRAS.478.4738N} {478, 4738}

\bibitem[\protect\citeauthoryear{{O'Riordan}, {Warren}  \&
  {Mortlock}}{{O'Riordan} et~al.}{2019}]{2019MNRAS.487.5143O}
{O'Riordan} C.~M.,  {Warren} S.~J.,   {Mortlock} D.~J.,  2019, \mn@doi [\mnras]
  {10.1093/mnras/stz1603}, \href
  {https://ui.adsabs.harvard.edu/abs/2019MNRAS.487.5143O} {487, 5143}

\bibitem[\protect\citeauthoryear{{Oort}}{{Oort}}{1932}]{1932BAN.....6..249O}
{Oort} J.~H.,  1932, \bain, \href
  {https://ui.adsabs.harvard.edu/abs/1932BAN.....6..249O} {6, 249}

\bibitem[\protect\citeauthoryear{{Osipkov}}{{Osipkov}}{1979}]{1979PAZh....5...77O}
{Osipkov} L.~P.,  1979, Pisma v Astronomicheskii Zhurnal, \href
  {https://ui.adsabs.harvard.edu/abs/1979PAZh....5...77O} {5, 77}

\bibitem[\protect\citeauthoryear{{Planck Collaboration}}{{Planck
  Collaboration}}{2016}]{2016A&A...594A..13P}
{Planck Collaboration} 2016, \mn@doi [\aap] {10.1051/0004-6361/201525830},
  \href {https://ui.adsabs.harvard.edu/abs/2016A&A...594A..13P} {594, A13}

\bibitem[\protect\citeauthoryear{{Plummer}}{{Plummer}}{1911}]{1911MNRAS..71..460P}
{Plummer} H.~C.,  1911, \mn@doi [\mnras] {10.1093/mnras/71.5.460}, \href
  {http://adsabs.harvard.edu/abs/1911MNRAS..71..460P} {71, 460}

\bibitem[\protect\citeauthoryear{{Prada}, {Klypin}, {Simonneau},
  {Betancort-Rijo}, {Patiri}, {Gottl{\"o}ber}  \& {Sanchez-Conde}}{{Prada}
  et~al.}{2006}]{2006ApJ...645.1001P}
{Prada} F.,  {Klypin} A.~A.,  {Simonneau} E.,  {Betancort-Rijo} J.,  {Patiri}
  S.,  {Gottl{\"o}ber} S.,   {Sanchez-Conde} M.~A.,  2006, \mn@doi [\apj]
  {10.1086/504456}, \href
  {https://ui.adsabs.harvard.edu/abs/2006ApJ...645.1001P} {645, 1001}

\bibitem[\protect\citeauthoryear{{Retana-Montenegro}, {van Hese}, {Gentile},
  {Baes}  \& {Frutos-Alfaro}}{{Retana-Montenegro}
  et~al.}{2012a}]{2012A&A...540A..70R}
{Retana-Montenegro} E.,  {van Hese} E.,  {Gentile} G.,  {Baes} M.,
  {Frutos-Alfaro} F.,  2012a, \mn@doi [\aap] {10.1051/0004-6361/201118543},
  \href {http://adsabs.harvard.edu/abs/2012A%26A...540A..70R} {540, A70}

\bibitem[\protect\citeauthoryear{{Retana-Montenegro}, {Frutos-Alfaro}  \&
  {Baes}}{{Retana-Montenegro} et~al.}{2012b}]{2012A&A...546A..32R}
{Retana-Montenegro} E.,  {Frutos-Alfaro} F.,   {Baes} M.,  2012b, \mn@doi
  [\aap] {10.1051/0004-6361/201219539}, \href
  {http://adsabs.harvard.edu/abs/2012A%26A...546A..32R} {546, A32}

\bibitem[\protect\citeauthoryear{{Richstone} \& {Tremaine}}{{Richstone} \&
  {Tremaine}}{1984}]{1984ApJ...286...27R}
{Richstone} D.~O.,  {Tremaine} S.,  1984, \mn@doi [\apj] {10.1086/162572},
  \href {https://ui.adsabs.harvard.edu/abs/1984ApJ...286...27R} {286, 27}

\bibitem[\protect\citeauthoryear{{Rood}, {Page}, {Kintner}  \& {King}}{{Rood}
  et~al.}{1972}]{1972ApJ...175..627R}
{Rood} H.~J.,  {Page} T.~L.,  {Kintner} E.~C.,   {King} I.~R.,  1972, \mn@doi
  [\apj] {10.1086/151585}, \href
  {https://ui.adsabs.harvard.edu/abs/1972ApJ...175..627R} {175, 627}

\bibitem[\protect\citeauthoryear{{Rubin}, {Ford}  \& {Thonnard}}{{Rubin}
  et~al.}{1978}]{1978ApJ...225L.107R}
{Rubin} V.~C.,  {Ford} W.~K. J.,   {Thonnard} N.,  1978, \mn@doi [\apjl]
  {10.1086/182804}, \href
  {https://ui.adsabs.harvard.edu/abs/1978ApJ...225L.107R} {225, L107}

\bibitem[\protect\citeauthoryear{{Schneider} \& {Sluse}}{{Schneider} \&
  {Sluse}}{2014}]{2014A&A...564A.103S}
{Schneider} P.,  {Sluse} D.,  2014, \mn@doi [\aap]
  {10.1051/0004-6361/201322106}, \href
  {https://ui.adsabs.harvard.edu/abs/2014A&A...564A.103S} {564, A103}

\bibitem[\protect\citeauthoryear{{Schwarzschild}}{{Schwarzschild}}{1979}]{1979ApJ...232..236S}
{Schwarzschild} M.,  1979, \mn@doi [\apj] {10.1086/157282}, \href
  {https://ui.adsabs.harvard.edu/abs/1979ApJ...232..236S} {232, 236}

\bibitem[\protect\citeauthoryear{{S\'ersic}}{{S\'ersic}}{1968}]{1968adga.book.....S}
{S\'ersic} J.~L.,  1968, {Atlas de Galaxias Australes}.
Observatorio Astronomico, Cordoba, Argentina

\bibitem[\protect\citeauthoryear{{Sonnenfeld}}{{Sonnenfeld}}{2018}]{2018MNRAS.474.4648S}
{Sonnenfeld} A.,  2018, \mn@doi [\mnras] {10.1093/mnras/stx3105}, \href
  {https://ui.adsabs.harvard.edu/abs/2018MNRAS.474.4648S} {474, 4648}

\bibitem[\protect\citeauthoryear{{Suyu}}{{Suyu}}{2012}]{2012MNRAS.426..868S}
{Suyu} S.~H.,  2012, \mn@doi [\mnras] {10.1111/j.1365-2966.2012.21661.x}, \href
  {https://ui.adsabs.harvard.edu/abs/2012MNRAS.426..868S} {426, 868}

\bibitem[\protect\citeauthoryear{{Swaters}, {Madore}, {van den Bosch}  \&
  {Balcells}}{{Swaters} et~al.}{2003}]{2003ApJ...583..732S}
{Swaters} R.~A.,  {Madore} B.~F.,  {van den Bosch} F.~C.,   {Balcells} M.,
  2003, \mn@doi [\apj] {10.1086/345426}, \href
  {https://ui.adsabs.harvard.edu/abs/2003ApJ...583..732S} {583, 732}

\bibitem[\protect\citeauthoryear{{Taylor} \& {Navarro}}{{Taylor} \&
  {Navarro}}{2001}]{2001ApJ...563..483T}
{Taylor} J.~E.,  {Navarro} J.~F.,  2001, \mn@doi [\apj] {10.1086/324031}, \href
  {https://ui.adsabs.harvard.edu/abs/2001ApJ...563..483T} {563, 483}

\bibitem[\protect\citeauthoryear{{Tessore} \& {Metcalf}}{{Tessore} \&
  {Metcalf}}{2015}]{2015A&A...580A..79T}
{Tessore} N.,  {Metcalf} R.~B.,  2015, \mn@doi [\aap]
  {10.1051/0004-6361/201526773}, \href
  {https://ui.adsabs.harvard.edu/abs/2015A&A...580A..79T} {580, A79}

\bibitem[\protect\citeauthoryear{{Tollet} et~al.,}{{Tollet}
  et~al.}{2016}]{2016MNRAS.456.3542T}
{Tollet} E.,  et~al., 2016, \mn@doi [\mnras] {10.1093/mnras/stv2856}, \href
  {https://ui.adsabs.harvard.edu/abs/2016MNRAS.456.3542T} {456, 3542}

\bibitem[\protect\citeauthoryear{{Tremaine}, {Richstone}, {Byun}, {Dressler},
  {Faber}, {Grillmair}, {Kormendy}  \& {Lauer}}{{Tremaine}
  et~al.}{1994}]{1994AJ....107..634T}
{Tremaine} S.,  {Richstone} D.~O.,  {Byun} Y.-I.,  {Dressler} A.,  {Faber}
  S.~M.,  {Grillmair} C.,  {Kormendy} J.,   {Lauer} T.~R.,  1994, \mn@doi [\aj]
  {10.1086/116883}, \href {http://adsabs.harvard.edu/abs/1994AJ....107..634T}
  {107, 634}

\bibitem[\protect\citeauthoryear{{Treu}}{{Treu}}{2010}]{2010ARA&A..48...87T}
{Treu} T.,  2010, \mn@doi [\araa] {10.1146/annurev-astro-081309-130924}, \href
  {https://ui.adsabs.harvard.edu/abs/2010ARA&A..48...87T} {48, 87}

\bibitem[\protect\citeauthoryear{{Van Hese}, {Baes}  \& {Dejonghe}}{{Van Hese}
  et~al.}{2009}]{2009ApJ...690.1280V}
{Van Hese} E.,  {Baes} M.,   {Dejonghe} H.,  2009, \mn@doi [\apj]
  {10.1088/0004-637X/690/2/1280}, \href
  {http://adsabs.harvard.edu/abs/2009ApJ...690.1280V} {690, 1280}

\bibitem[\protect\citeauthoryear{{Veltmann}}{{Veltmann}}{1979}]{1979AZh....56..976V}
{Veltmann} U.~I.~K.,  1979, \azh, \href
  {http://adsabs.harvard.edu/abs/1979AZh....56..976V} {56, 976}

\bibitem[\protect\citeauthoryear{{Wang}, {Chen}  \& {Li}}{{Wang}
  et~al.}{2017}]{2017MNRAS.471..523W}
{Wang} L.,  {Chen} D.-M.,   {Li} R.,  2017, \mn@doi [\mnras]
  {10.1093/mnras/stx1617}, \href
  {https://ui.adsabs.harvard.edu/abs/2017MNRAS.471..523W} {471, 523}

\bibitem[\protect\citeauthoryear{{Wertz} \& {Surdej}}{{Wertz} \&
  {Surdej}}{2014}]{2014MNRAS.437.1051W}
{Wertz} O.,  {Surdej} J.,  2014, \mn@doi [\mnras] {10.1093/mnras/stt1815},
  \href {https://ui.adsabs.harvard.edu/abs/2014MNRAS.437.1051W} {437, 1051}

\bibitem[\protect\citeauthoryear{{Wojtak}, {Gottl{\"o}ber}  \&
  {Klypin}}{{Wojtak} et~al.}{2013}]{2013MNRAS.434.1576W}
{Wojtak} R.,  {Gottl{\"o}ber} S.,   {Klypin} A.,  2013, \mn@doi [\mnras]
  {10.1093/mnras/stt1113}, \href
  {https://ui.adsabs.harvard.edu/abs/2013MNRAS.434.1576W} {434, 1576}

\bibitem[\protect\citeauthoryear{{Wright} \& {Brainerd}}{{Wright} \&
  {Brainerd}}{2000}]{2000ApJ...534...34W}
{Wright} C.~O.,  {Brainerd} T.~G.,  2000, \mn@doi [\apj] {10.1086/308744},
  \href {https://ui.adsabs.harvard.edu/abs/2000ApJ...534...34W} {534, 34}

\bibitem[\protect\citeauthoryear{{Xu}, {Sluse}, {Schneider}, {Springel},
  {Vogelsberger}, {Nelson}  \& {Hernquist}}{{Xu}
  et~al.}{2016}]{2016MNRAS.456..739X}
{Xu} D.,  {Sluse} D.,  {Schneider} P.,  {Springel} V.,  {Vogelsberger} M.,
  {Nelson} D.,   {Hernquist} L.,  2016, \mn@doi [\mnras]
  {10.1093/mnras/stv2708}, \href
  {https://ui.adsabs.harvard.edu/abs/2016MNRAS.456..739X} {456, 739}

\bibitem[\protect\citeauthoryear{{Zhao}}{{Zhao}}{1996}]{1996MNRAS.278..488Z}
{Zhao} H.,  1996, \mn@doi [\mnras] {10.1093/mnras/278.2.488}, \href
  {http://adsabs.harvard.edu/abs/1996MNRAS.278..488Z} {278, 488}

\bibitem[\protect\citeauthoryear{{Zhao}}{{Zhao}}{1997}]{1997MNRAS.287..525Z}
{Zhao} H.,  1997, \mn@doi [\mnras] {10.1093/mnras/287.3.525}, \href
  {https://ui.adsabs.harvard.edu/abs/1997MNRAS.287..525Z} {287, 525}

\bibitem[\protect\citeauthoryear{{Zwicky}}{{Zwicky}}{1933}]{1933AcHPh...6..110Z}
{Zwicky} F.,  1933, Helvetica Physica Acta, \href
  {https://ui.adsabs.harvard.edu/abs/1933AcHPh...6..110Z} {6, 110}

\bibitem[\protect\citeauthoryear{{de Blok}, {McGaugh}, {Bosma}  \& {Rubin}}{{de
  Blok} et~al.}{2001}]{2001ApJ...552L..23D}
{de Blok} W.~J.~G.,  {McGaugh} S.~S.,  {Bosma} A.,   {Rubin} V.~C.,  2001,
  \mn@doi [\apjl] {10.1086/320262}, \href
  {https://ui.adsabs.harvard.edu/abs/2001ApJ...552L..23D} {552, L23}

\bibitem[\protect\citeauthoryear{{de Blok}, {Walter}, {Brinks}, {Trachternach},
  {Oh}  \& {Kennicutt}}{{de Blok} et~al.}{2008}]{2008AJ....136.2648D}
{de Blok} W.~J.~G.,  {Walter} F.,  {Brinks} E.,  {Trachternach} C.,  {Oh}
  S.~H.,   {Kennicutt} R.~C. J.,  2008, \mn@doi [\aj]
  {10.1088/0004-6256/136/6/2648}, \href
  {https://ui.adsabs.harvard.edu/abs/2008AJ....136.2648D} {136, 2648}

\bibitem[\protect\citeauthoryear{{de Zeeuw}}{{de
  Zeeuw}}{1985}]{1985MNRAS.216..273D}
{de Zeeuw} T.,  1985, \mn@doi [\mnras] {10.1093/mnras/216.2.273}, \href
  {http://adsabs.harvard.edu/abs/1985MNRAS.216..273D} {216, 273}

\makeatother
\end{thebibliography}

\appendix
\onecolumn
\section{Analytical expressions for the BPL model with $(\beta,\gamma)=(4,1)$}
\label{AppendixA.sec}

In the main body of this paper we have used the BPL model with $(\beta,\gamma)=(4,1)$ as our fiducial model. The main reason why we chose this particular model is that the subfamily of Zhao models with $(\beta,\gamma)=(4,1)$ contains the Hernquist model as a special case, and can hence be regarded as a generalisation of this often used and well-studied model \citep[e.g.,][]{1990ApJ...356..359H, 2002A&A...393..485B, 2007A&A...471..419B}. An additional benefit is that, for this BPL model, most of the dynamical properties, including the the distribution function and the differential energy distribution, can be calculated analytically.

For the basic properties we can immediately find the analytical expressions by setting $\beta=4$ and $\gamma=1$ in the expressions (\ref{BPL-rho}), (\ref{BPL-M}), (\ref{BPL-Psi}) and (\ref{BPL-sigma2}),
\begin{gather}
\rho_{\text{in}} = \frac{1}{r},
\\
\rho_{\text{out}} = \frac{1}{r^4},
\\
M_{\text{in}} = 2\pi\,r^2,
\\
M_{\text{out}} = 4\pi\left(\frac32-\frac{1}{r}\right),
\\
\Psi_{\text{in}} =2\pi\left(3-r\right),
\\
\Psi_{\text{out}} = \frac{2\pi}{r}\left(3-\frac{1}{r}\right),
\\
\sigma^2_{\text{in}} = 2\pi\,r\left(\frac{4}{15}-\ln r\right),
\\
\sigma^2_{\text{out}} = \frac{2\pi}{r}\left(\frac35-\frac{1}{3r}\right).
\end{gather}
The model has a finite total mass and a finite potential well, with
\begin{gather}
M = 6\pi,
\\
\Psib = 4\pi,
\\
\Psi_0 = 6\pi.
\end{gather}
For the distribution function and the density of states function, we need to substitute the potential in the expressions~(\ref{BPL-df}) and (\ref{gE2}), and evaluate the resulting integrals. For the inner part, corresponding to $\Psib<\calE\leqslant\Psi_0$ we find
\begin{equation}
f_{\text{in}}(\calE) = \frac{1}{16\,\pi^{7/2}}
\left[
\frac{32y^4+20y^3-64y^2+15y-9}{(4y-3)\,y^2\sqrt{1-y}}
+
\frac{4\,(8y+3)\sqrt{3-y}}{4y-3}
+
\frac{3\arccos(\sqrt{y})}{y^{5/2}}
+
18\ln\left(\frac{3-2\sqrt{3-y}}{1-2\sqrt{1-y}}\right)\right]
\end{equation}
and
\begin{equation}
g_{\text{in}}(\calE) = \frac{512\,\pi^{5/2}}{105}\,y^{7/2}
\end{equation}
with
\begin{equation}
y = 3-\frac{\calE}{2\pi}.
\end{equation}
For the outer part, corresponding to binding energies $\calE<\Psib$, we have
\begin{equation}
f_{\text{out}}(\calE) 
= \frac{1}{8\pi^{7/2}}
\left[\frac{2\,(8-24x+27x^2)\sqrt{3x-1}}{x\,(3x-2)^2}
+9\ln\left(\frac{3x-2\sqrt{3x-1}}{3x-2}\right)\right]
\end{equation}
and
\begin{multline}
g_{\text{out}}(\calE) = \frac{\pi^{5/2}}{105}
\left[ 
\frac{512\,(3x^2-3x+1)^{7/2}}{x^7}
+
\frac{140\,(27x^4-6x^3-70x^2+48x-8)\sqrt{(x-1)\,(2x-1)}}{x\,(3x-1)^2}
\right.
\\
+
\left.
\frac{64\,(246x^6-729x^5+927x^4-672x^3+292x^2-72x+8)\sqrt{1-3x+2x^2}}{x^7}
\right.
\\
\left.
+ \frac{630\,(3x-2)^2\,x^3}{(3x-1)^{5/2}}
\left(\frac{\pi}{2}+\arctan\left(\frac{3x^2-6x+2}{2\sqrt{(x-1)\,(2x-1)\,(3x-1)}}\right)\right)
\right]
\end{multline}
with
\begin{equation}
x = \frac{3\pi}{\calE}\left(1+\sqrt{1-\dfrac{2\calE}{9\pi}}\right).
\end{equation}
At small binding energies ($\calE\to0$), the distribution function, the density-of-states function, and the differential energy distribution behave as
\begin{gather}
f(\calE) 
= \frac{\sqrt2}{405\,\pi^6}\,\calE^{5/2} + \cdots,
\\
g(\calE) 
= 216\sqrt2\,\pi^6\,\calE^{-5/2} + \cdots,
\\
\calN(\calE) 
= \frac{15}{16} + \frac{32}{315\,\pi}\,\calE + \cdots,
\end{gather}
whereas for $\calE\to\Psi_0$ we have the expansions
\begin{gather}
f(\calE) 
= \frac{3}{4\sqrt2} \left(\Psi_0-\calE\right)^{-5/2} + \cdots,
\\
g(\calE) 
= \frac{32\sqrt2}{105\,\pi} \left(\Psi_0-\calE\right)^{7/2},
\\
\calN(\calE) 
= \frac{8}{35\,\pi} \left(\Psi_0-\calE\right) + \cdots.
\end{gather}
The leading terms in these expansions have the same order as those of the Hernquist model, as expected. The explicit expression for the distribution function also allows an investigation of the behaviour in the neighbourhood of $\Psib$. For $\calE$ approaching $\Psib$ from the low binding energy side, we find 
\begin{equation}
f(\calE) 
=
\frac{22\sqrt2+9\ln(3-2\sqrt2)}{8\,\pi^{7/2}}  
- \frac{4\sqrt2}{\pi^{9/2}}\left(\Psib-\calE\right)
+ \cdots.
\end{equation}
Approaching $\Psib$ from the high binding energy side, we find the expansion 
\begin{equation}
f(\calE) 
=
-\frac{3}{4\sqrt2\,\pi^3} \frac{1}{\sqrt{\calE-\Psib}}
+
\frac{22\sqrt2+9\ln(3-2\sqrt2)}{8\,\pi^{7/2}}  
-
\frac{9}{2\sqrt2\,\pi^4} \sqrt{\calE-\Psib}
+ \cdots.
\end{equation}
This clearly demonstrates the discontinuity of the distribution function at $\calE=\Psib$, with a finite value at the low binding energy side and an inverse square-root divergence to $-\infty$ on the high binding energy side. 

\label{lastpage}
\end{document}